\documentclass[12pt]{article}
\usepackage{amssymb}
\usepackage{graphicx}
\begin{document}
\newcommand{\beq}{\begin{equation}}
\newcommand{\eeq}{\end{equation}}
\newcommand{\beqa}{\begin{eqnarray}}
\newcommand{\eeqa}{\end{eqnarray}}
\newcommand{\beqar}{\begin{eqnarray*}}
\newcommand{\eeqar}{\end{eqnarray*}}
\newcommand{\al}{\alpha}
\newcommand{\be}{\beta}
\newcommand{\del}{\delta}
\newcommand{\D}{\Delta}
\newcommand{\eps}{\epsilon}
\newcommand{\ga}{\gamma}
\newcommand{\Ga}{\Gamma}
\newcommand{\ka}{\kappa}
\newcommand{\nn}{\nonumber}
\newcommand{\inn}{\!\cdot\!}
\newcommand{\h}{\eta}
\newcommand{\ii}{\iota}
\newcommand{\kk}{\varphi}
\newcommand\F{{}_3F_2}
\newcommand{\la}{\lambda}
\newcommand{\La}{\Lambda}
\newcommand{\na}{\prt}
\newcommand{\Om}{\Omega}
\newcommand{\om}{\omega}
\newcommand{\p}{\phi}
\newcommand{\sig}{\sigma}
\renewcommand{\t}{\theta}
\newcommand{\z}{\zeta}
\newcommand{\ssc}{\scriptscriptstyle}
\newcommand{\eg}{{\it e.g.,}\ }
\newcommand{\ie}{{\it i.e.,}\ }
\newcommand{\labell}[1]{\label{#1}} 
\newcommand{\reef}[1]{(\ref{#1})}
\newcommand\prt{\partial}
\newcommand\veps{\varepsilon}
\newcommand{\pol}{\varepsilon}
\newcommand\vp{\varphi}
\newcommand\ls{\ell_s}
\newcommand\cF{{\cal F}}
\newcommand\cA{{\cal A}}
\newcommand\cS{{\cal S}}
\newcommand\cT{{\cal T}}
\newcommand\cV{{\cal V}}
\newcommand\cL{{\cal L}}
\newcommand\cM{{\cal M}}
\newcommand\cN{{\cal N}}
\newcommand\cG{{\cal G}}
\newcommand\cK{{\cal K}}
\newcommand\cH{{\cal H}}
\newcommand\cI{{\cal I}}
\newcommand\cJ{{\cal J}}
\newcommand\cl{{\iota}}
\newcommand\cP{{\cal P}}
\newcommand\cQ{{\cal Q}}
\newcommand\cg{{\it g}}
\newcommand\cR{{\cal R}}
\newcommand\cB{{\cal B}}
\newcommand\cO{{\cal O}}
\newcommand\tcO{{\tilde {{\cal O}}}}
\newcommand\bz{\bar{z}}
\newcommand\bc{\bar{c}}
\newcommand\bw{\bar{w}}
\newcommand\bX{\bar{X}}
\newcommand\bK{\bar{K}}
\newcommand\bA{\bar{A}}
\newcommand\bZ{\bar{Z}}
\newcommand\bxi{\bar{\xi}}
\newcommand\bphi{\bar{\phi}}
\newcommand\bpsi{\bar{\psi}}
\newcommand\bprt{\bar{\prt}}
\newcommand\bet{\bar{\eta}}
\newcommand\btau{\bar{\tau}}
\newcommand\hF{\hat{F}}
\newcommand\hA{\hat{A}}
\newcommand\hT{\hat{T}}
\newcommand\htau{\hat{\tau}}
\newcommand\hD{\hat{D}}
\newcommand\hf{\hat{f}}
\newcommand\hg{\hat{g}}
\newcommand\hp{\hat{\phi}}
\newcommand\hi{\hat{i}}
\newcommand\ha{\hat{a}}
\newcommand\hb{\hat{b}}
\newcommand\hQ{\hat{Q}}
\newcommand\hP{\hat{\Phi}}
\newcommand\hS{\hat{S}}
\newcommand\hX{\hat{X}}
\newcommand\tL{\tilde{\cal L}}
\newcommand\hL{\hat{\cal L}}
\newcommand\tG{{\tilde G}}
\newcommand\tg{{\tilde g}}
\newcommand\tphi{{\widetilde \phi}}
\newcommand\tPhi{{\widetilde \Phi}}
\newcommand\te{{\tilde e}}
\newcommand\tk{{\tilde k}}
\newcommand\tf{{\tilde f}}
\newcommand\ta{{\tilde a}}
\newcommand\tb{{\tilde b}}
\newcommand\tc{{\tilde c}}
\newcommand\td{{\tilde d}}
 \newcommand\tR{{\tilde R}}
\newcommand\teta{{\tilde \eta}}
\newcommand\tF{{\widetilde F}}
\newcommand\tK{{\widetilde K}}
\newcommand\tE{{\widetilde E}}
\newcommand\tpsi{{\tilde \psi}}
\newcommand\tX{{\widetilde X}}
\newcommand\tD{{\widetilde D}}
\newcommand\tO{{\widetilde O}}
\newcommand\tS{{\tilde S}}
\newcommand\tB{{\widetilde B}}
\newcommand\tA{{\widetilde A}}
\newcommand\tT{{\widetilde T}}
\newcommand\tC{{\widetilde C}}
\newcommand\tV{{\widetilde V}}
\newcommand\thF{{\widetilde {\hat {F}}}}
\newcommand\Tr{{\rm Tr}}
\newcommand\tr{{\rm tr}}
\newcommand\STr{{\rm STr}}
\newcommand\hR{\hat{R}}
\newcommand\M[2]{M^{#1}{}_{#2}}

\newcommand\bS{\textbf{ S}}
\newcommand\bI{\textbf{ I}}
\newcommand\bJ{\textbf{ J}}

\begin{titlepage}
\begin{center}

\vskip 2 cm
{\LARGE \bf   Couplings of order six in the gauge field strength and the second
    fundamental form\\ on a D$_p$-brane at order $\alpha {'^2}$
 }\\
\vskip 1.25 cm
  Mohammad R. Garousi\footnote{garousi@um.ac.ir}
and Saman Karimi\footnote{karimi.saman@mail.um.ac.ir}

\vskip 1 cm
{{\it Department of Physics, Faculty of Science, Ferdowsi University of Mashhad\\}{\it P.O. Box 1436, Mashhad, Iran}\\}
\vskip .1 cm
 \end{center}

\begin{abstract}

Using the assumption that the independent gauge invariant couplings on the world-volume of the non-perturbative objects in the string theory are independent of the background, we find the four and the six gauge field strength and/or the second fundamental form couplings on the world volume of a D$_p$-brane in the superstring theory at order $\alpha'^2$ in the normalization that $F$ is dimensionless. We have found them by considering the particular background which has one circle and by imposing the corresponding T-duality constraint on the independent couplings.
In particular, we find that there are 12+146 independent gauge invariant couplings at this order, and the T-duality constraint can fix 150 of them. We show that these couplings are fully consistent with the partial results  in the literature. This comparison also fixes the remaining 8 couplings.  

\end{abstract}

\end{titlepage}

\section{Introduction}

The critical string theory is a quantum theory of gravity which reproduces the Einstein theory of general relativity at the low energy.  As in the Einstein theory, one expects the string theory and its non-perturbative objects at the critical dimension   to be background independent. In the low energy effective action, the background independence  means the coefficients of the independent  gauge invariant couplings at each order of $\alpha'$ should be   independent of the background. If one could fix these coefficients in a particular background in which the effective action has some symmetries, then that coefficients  would be valid  for any other background which may have no symmetry. 

The independent couplings at a given order of $\alpha'$  are  given as all gauge invariant and covariant couplings at that order  modulo  the field redefinitions, the total derivative terms and the Bianchi identities.  The numbers of independent couplings  in the bosonic string theory involving the metric, dilaton and the $B$-field at orders $\alpha',\alpha'^2,\alpha'^3$ are $8,60,872$, respectively \cite{Metsaev:1987zx,Garousi:2019cdn,Garousi:2020mqn}. The number of independent world-volume couplings of O$_p$-plane in the superstring theory at order $\alpha'^2$ involving only NS-NS fields is 48 \cite{Akou:2020mxx}, and involving linear R-R field and the NS-NS fields  is 77 \cite{Mashhadi:2020mzf}.   The background independent  coefficients  of all these couplings are  fixed  when one considers a particular background which includes one  circle, and uses the corresponding T-duality constraints \cite{Garousi:2019wgz,Garousi:2019mca,Razaghian:2018svg,Garousi:2020gio,Garousi:2020lof}. One may also use the background independence assumption to find the boundary couplings in the case that the background has boundary \cite{Garousi:2019xlf,Akou:2020mxx,Garousi:2022ovo}.

The world-volume gauge invariant couplings of a non-perturbative D$_p$-brane involving open string massless gauge-fields/transverse-scalars at long wavelength limit is given by  the  Dirac-Born-Infeld (DBI) action \cite{Leigh:1989jq,Bachas:1995kx}
\beqa
S_p&=&-T_p\int d^{p+1}\sigma\sqrt{-\det(\tG_{ab}+  F_{ab})} \labell{a1}
\eeqa
 where $ {T_p} $ is tension of $ {D_p} $-brane, $ {{F_{ab}}} $ is  field strength of the gauge field $ {A_a} $  and $ {{{\tilde G}_{ab}}} $ is the pull-back of the bulk  metric onto the world-volume\footnote{Our index convention is that the Greek letters  $(\mu,\nu,\cdots)$ are  the indices of the space-time coordinates, the Latin letters $(a,d,c,\cdots)$ are the world-volume indices and the Latin letters $(i,j,k,\cdots)$ are the transverse indices. }, \ie
\beqa
\tG_{ab}&=&\frac{\prt X^{\mu}(\sigma)}{\prt\sigma^a}\frac{\prt X^{\nu}(\sigma)}{\prt\sigma^b}\eta_{\mu\nu}\,\equiv \, \prt_a X^\mu\prt_b X^\nu \eta_{\mu\nu}\labell{pull}
\eeqa
where $ {{X^\mu }(\sigma)} $ is the spacetime coordinate which specifies the D$_p$-brane in the spacetime,  and $ {\eta _{\mu \nu }} $ is  the spacetime metric which for simplicity we choose it to be  the Minkowski metric. We have also chosen the B-field  to be zero and the dilaton to be a constant.  We have normalized the gauge field $A_a$ to have the same dimension as the world sheet field $X^\mu$. With this normalization, the above action is at the leading order of $\alpha'$.  The above action includes all even-power of the gauge field strength $F_{ab}$.   The transverse scalar fields $\Phi^i$ appear in the static gauge where $X^a=\sigma^a, X^i=\Phi^i(\sigma)$. In the static gauge and for  $\Phi^i=0$, the DBI action reduces to the Born-Infeld action. The $\alpha'$ corrections to the Born-Infeld action  have been studied in \cite{Abouelsaood:1986gd,Tseytlin:1987ww,Andreev:1988cb, Wyllard:2000qe, Andreev:2001xx, Wyllard:2001ye}.

 In the superstring theory, the first correction to the DBI  action is at order  $\alpha {'^2}$ which involves some contractions of  the second fundamental form $\Omega_{ab}{}^\mu$, \ie
 \beqa
\Omega _{ab}^{\,\,\,\,\,\,\,\mu } = {D_a}{\partial _b}{X^\mu }\,, \labell{a5}
\eeqa
 the gauge field strength $F_{ab}$ and their covariant derivatives, \eg
\begin{eqnarray}
  {D_a}{F_{bc}} = {\partial _a}{F_{bc}} - \tilde \Gamma _{ab}^{\,\,\,\,\,d}{F_{dc}} - \tilde \Gamma _{ac}^{\,\,\,\,\,d}{F_{bd}}\,,
   \labell{a3}
\end{eqnarray}
where  the Levi-Civita  connection  $ \tilde \Gamma _{ab}^{\,\,\,\,\,c} $  is made of the pull-back metric   \reef{pull}.  The world volume indices of these gauge invariant tensors  are contracted with the inverse of the pull-back metric  $\tG^{ab}$, and the spacetime index in the second fundamental form are contracted with the spacetime metric $\eta_{\mu\nu}$.  Even though the $\alpha'^2$-order of the couplings,  constrains the independent couplings to have at most the first derivative of $\Omega$ and the second derivative of $F$, however, there are infinite tower of the  gauge field strength, without derivative on it, in the couplings. Hence, for simplicity we consider only the couplings at order $\alpha'^2$ which involve at most six gauge field strengths and/or the second fundamental form. Using the background independence assumption, we are going to find such couplings in this paper. That is, we first find the independent gauge invariant couplings and then consider a particular background which has one circle. For this background, the couplings should satisfy the T-duality constraint \cite{ Robbins:2014ara,Garousi:2017fbe},  \ie  the T-duality transformation of the world-volume reduction of the independent covariant couplings must be the same as the transverse reduction of the couplings, up to some total derivative terms and field redefinitions in the base space. This constraint  may fix   the coefficients of the independent couplings. This method has been used in \cite{Karimi:2018vaf} to find  the corrections to the DBI action in the bosonic string theory at order $\alpha'$ which involve at  most eight gauge field strengths and/or the second fundamental forms. The covariant approach has been used in \cite{Karimi:2018vaf} to find the independent couplings, however, the  T-duality constraint  has been used  in the static gauge. In this paper, we are going to use the covariant approach for finding the independent couplings as well as  for  imposing the T-duality constraint.

The outline of the paper is as follows: In section 2, we find all  independent covariant couplings at order $\alpha'^2$ which involve at most six gauge fields and/or the second fundamental forms. We find there is no independent couplings at the level of  two fields, there are 12 independent couplings at the level of four fields, and there are 146 couplings at the six-field level.   The coefficients of these couplings are independent of the backgrounds in which  the D$_p$-brane are placed.  To fix these 158 background independent  coefficients, in section 3, we consider a background which includes a circle. Then the independent couplings must satisfy the  T-duality constraint.  We find that the T-duality constraint fixes the 12 parameters of the four-field couplings up to five parameters. They are consistent with the couplings that are found in the literature by the S-matrix method. We use this comparison to fix the remaining 5 parameters. We then find  that the T-duality constraint fixes 145 parameters of the six-field couplings. We show that the couplings which involve only the gauge field are consistent with the all-gauge-field couplings that are found by Wyllard in \cite{Wyllard:2000qe}. We also fix the remaining 3 parameters by this comparison. In section 4, we extend the all-gauge-field couplings found in \cite{Wyllard:2000qe} to covariant form and found their  corresponding four-field and six-field couplings involving the second fundamental form. In section 5, we briefly discuss our results.

\section{Independent couplings}

In this section we are going to find the independent couplings at order $\alpha'^2$ which involve at most six gauge-field and/or the second fundamental form. We apply the method used in \cite{Garousi:2019cdn} to find the independent couplings.   The independent couplings   are  all gauge invariant couplings modulo  the field redefinitions, the total derivative terms, the identities corresponding to the derivative of the second fundamental form, the Bianchi identity corresponding to the gauge field
\beqa
\prt_{[a}F_{bc]}=0\labell{bian}
\eeqa
and the following identity involving  the second fundamental form and $\prt _a X^\mu$:
\beqa
{\Omega _{ab}{}^\mu}{\partial _c}{X^\nu } \eta_{\mu\nu}= 0 \labell{Identity}
\eeqa
The above identity can easily be verified by using \reef{a5} and writing the covariant derivative in terms of partial derivative and the  Levi-Civita connection, and then writing the connection in terms of the pull-back metric \reef{pull}.  Using the above identity, one finds that there is a scheme in which  $\prt X$ can appear only through the pull-back metric \reef{pull} and its inverse. For example the coupling  $D\Omega\prt X$ can be written as $-\Omega\Omega$ which can easily be verified by taking covariant derivative of the above identity.  Hence, we use the scheme in which  the couplings  involve only the contractions of $F, \Omega$ and their covariant derivatives, \ie
\beqa
S'&=&-\frac{(2\pi\alpha')^2}{96} T_p\int d^{p+1}\sigma \, \sqrt { - \det \tilde G_{ab}}\,\cL'(F,DF,\cdots, \Omega, D\Omega,\cdots)\labell{s1}
\eeqa
In principle,  one can construct all contractions  of the gauge-field strength and/or the second-fundamental form. We call the coefficients of these couplings $a_1',a_2',\cdots$.  However, they are not independent couplings.

To remove the total derivative terms from the gauge invariant couplings in \reef{s1}, we first construct a vector $\cI^a$ at order $\alpha'^{3/2}$ from  $F, \Omega$ and their covariant derivatives with arbitrary coefficients $z_1,z_2,\cdots$. Then one is free to add the following total derivative term to \reef{s1}:
\beqa
\cJ&=&-\alpha'^2 T_p\int d^{p+1}\sigma\, \sqrt { - \det \tilde G_{ab}} D_a\cI^a\labell{tot}
\eeqa
The  total derivative terms  may remove some of the structures in \reef{s1}
 completely, \eg $F DDDDF$ or $\Omega DD\Omega$, and may also  remove only some of the couplings in a particular structure in \reef{s1}. Hence, in writing the couplings in \reef{s1}, we do not include the structures that are removed completely by the total derivative terms.

 One is also free to change the field variables as\footnote{One may also consider the change of variables at order $\alpha'^{1/2}$ and consider the second perturbation of the DBI action which also produces couplings at order $\alpha'^2$. However, such field redefinition would also produce at the linear order, the couplings at order $\alpha'$ which is  in conflict with the fact that there is no world volume couplings in the superstring theory at order $\alpha'$. Hence, there should be no such field redefinition. }
\beqa
A_a &\rightarrow& A_a+\alpha'^{3/2}\delta A_a, \nonumber\\
{X^\mu } &\rightarrow& {X^\mu } + \alpha'^{3/2}\delta {X^\mu }
\labell{b40}
\eeqa
where the tensors $\delta A_a$ and $\delta {X^\mu }$ are all contractions of $F, \Omega$ and their covariant derivatives at order $\alpha'^{3/2}$ with arbitrary coefficients $y_1,y_2,\cdots$. If one replaces this field redefinition into the leading order action \reef{a1}, it would produces the following couplings at order $\alpha'^2$:
\beqa
\cK&=&-\alpha'^2T_p\int d^{p+1}\sigma\,
 \sqrt { - \det \tilde G_{ab}}  \Bigg[
-{D_a}{F^{ab}}\delta {A_b} - {\tilde G^{ab}}{\Omega _{ab}}^\nu \delta {X^\mu }{\eta _{\mu \nu }} +  \cdots \Bigg ]\labell{fred}
\eeqa
where dots represent the terms which involve all higher orders of $F$ resulting from the linear perturbation of the leading order action \reef{a1} around \reef{b40}. If one uses the arbitrary parameters in  $\delta A_a$ and $\delta {X^\mu }$ to remove all couplings in \reef{s1} which have $D_aF^{ab}$ and $G^{ab}{\Omega _{ab}}^\nu$, then there would be no residual arbitrary parameters in $\delta A_a$ and $\delta {X^\mu }$ to remove any  couplings in \reef{s1} that have the same structure as the couplings in the dots above. 
Therefore,  in the scheme that  the field redefinitions removes the  couplings that have  $D_aF^{ab}$ or $G^{ab}{\Omega _{ab}}^\nu$, one must  ignore the dots above.

If one adds $\cJ, \cK$ to the action \reef{s1}, they change only the coefficients of the gauge invariant couplings $a_1',a_2',\cdots$, \ie
\beqa
S'+\cJ+\cK=S\labell{SS}
\eeqa
where $S$ is the same action as \reef{s1} in which the coefficients of the gauge invariant couplings are changed to $a_1,a_2,\cdots$. One can write the above equation as
\beqa
\Delta S+\cJ+\cK&=&0\labell{SJK}
\eeqa
where $\Delta S$ is the same as \reef{s1} in which the  coefficients of the gauge invariant couplings are $\delta a_1,\delta a_2,\cdots$ where $\delta a_i=a_i'-a_i$.  If one solves the above equation, one would find some relations between only $\delta a_1,\delta a_2,\cdots$. The number of these relations represents the number of  couplings which are invariant under the field redefinitions and the total derivative terms.

However, to solve the equation \reef{SJK}, one has to impose the Bianchi identity \reef{bian} and the identities corresponding to the derivative of the second fundamental form, to write \reef{SJK} in terms of independent couplings. To impose the latter identities automatically, one can write the covariant derivatives in terms of partial derivatives and   the  Levi-Civita connection. Moreover, one can go to the local frame in which the  Levi-Civita connection is zero but its derivatives are not zero. Then, one can  write the derivatives of the connection in terms of the pull-back metric \reef{pull}. In the resulting expression,  then one has to replace  the two $\prt X$ in which their spacetime index are contracted with each other, \ie $\prt_a X^\mu\prt_b X^\nu\eta_{\mu\nu}$,  by the pull-back metric \reef{pull}.   To impose the Bianchi identity \reef{bian}, we write the terms that have partial derivative of the gauge field strength in terms of the gauge potential, \eg $\prt_a F_{bc}=\prt_a\prt_b A_c-\prt_a\prt_c A_b$. The resulting terms have non-covariant expressions $F$, $\prt\prt A$, $\prt\prt\prt A,\cdots$, and $\prt X$, $\prt\prt X$, $\prt\prt\prt X, \cdots$. The world-volume indices are contracted with the inverse of the pull-back metric \reef{pull}  and the spacetime indices are contracted with $\eta_{\mu\nu}$. In other words, the equation \reef{SJK} is  written in the local frame in terms of non-covariant but independent terms. The coefficients of the independent terms  which involves $\delta a_1,\delta a_2,\cdots, z_1,z_2,\cdots, y_1,y_2,\cdots$, must be zero. The solution of the  resulting linear algebraic equations, gives   $z_1,z_2,\cdots, z_n, y_1,y_2,\cdots, y_m$ in terms of   $z_{n+1},z_{n+2},\cdots,  y_{m+1},y_{m+2},\cdots$  and $\delta a_1,\delta a_2,\cdots$ in which we are not interested.   The solution also gives some  relations between only $\delta a_1,\delta a_2,\cdots$ in which we are interested. The number of the latter relations gives the number of independent couplings in \reef{SJK}.

Since there can be any number of gauge field strength $F_{ab}$ in the couplings at any order of $\alpha'$, there  are infinite number of independent couplings at each order of $\alpha'$.  Hence, we have to classify the independent couplings in sub-structures  in which their couplings are independent.  In our choice for the field redefinition that removes all terms that involve  $D_aF^{ab}$ and $G^{ab}{\Omega _{ab}}^\nu$, the field redefinition does not relate the terms which have different number of the gauge fields, to each other. The total derivative terms and the Bianchi identities do not relate the couplings with different number of gauge fields either. Hence, in our choice for the field redefinition, the number of independent couplings at each level  of gauge field is fixed. Moreover, the couplings that involve only $F,\Omega$ modulo the trace of $\Omega$, are not related to the other couplings by the field redefinitions, by the total derivative terms and by the Bianchi identity. Hence, we choose all such couplings at each level of gauge field as independent couplings. We use the above prescription to find all other independent couplings  at each level of gauge field.

When $X^\mu$ is constant, \ie $\Omega=0$, the independent couplings involve only the gauge field strength $F_{ab}$ and its partial derivatives.  The above prescription can be used to find the independent couplings in this case. In the case that $X^\mu$ is not constant, \ie $\Omega\neq 0$,  there is a scheme in which the independent couplings classify into two sets of couplings. One set of couplings is the same as the set of independent couplings in the case that $X^\mu$ is constant. The second set of couplings is the independent couplings which become zero when $X^\mu$ is constant. It has been shown in \cite{Garousi:2019cdn} that in fact there is such scheme for the independent couplings of the bosonic string theory for metric, B-field and dilaton at order $\alpha'^2$. In particular, it has been shown in \cite{Garousi:2019cdn} that there are 60 independent couplings at this order. In one particular scheme, the couplings have been written as two sets. One set which has 20 couplings, includes  the dilaton only as the overall factor $e^{-2\phi}$, and another set which has 40 couplings, includes the derivative of the dilaton. In this scheme, when the dilaton is constant, the couplings reduce to 20 couplings which are the independent couplings when dilaton is constant \cite{Jones:1988hk}. It has been shown in \cite{Garousi:2019cdn}, that there is also a scheme in which  the dilaton appears as overall factor in all  60 independent couplings. In this scheme, when the dilaton is constant, the number of couplings does  not change, however, the 60 couplings are not independent any more when dilaton is constant. In this paper we are going to use the scheme in which the independent couplings are such that when $X^\mu$ is constant, they reduce to the independent couplings of only gauge field.

We begin with the couplings that have zero  gauge field at order $\alpha'^2$. There are 4 couplings involving $\Omega\Omega\Omega\Omega$ modulo the trace of $\Omega$.  Apart from this structure, the Lagrangian in \reef{s1} has one structure as
\beqa
\cL'& \sim & D\Omega D\Omega
\labell{Action3}
\eeqa
Using the package xAct \cite{Nutma:2013zea}, one finds there are  5 couplings in the above structure. The vector in the total derivative \reef{tot} has one structure as
 \beqa
\cI &\sim &\Omega D\Omega
\labell{TOT2}
\eeqa
The field redefinitions $\delta A_a$ has no structure at this level and $\delta X^\mu$  has one structure as
\beqa
\delta {X^\mu } &\sim& DD\Omega
  \labell{b60b}
\eeqa
Using the package xAct, one can construct all possible contractions  in \reef{TOT2} and \reef{b60b}. Then we replace them in \reef{SJK} and go to the local frame to write the equation \reef{SJK} in terms of the independent structures. However, the coefficients of all resulting  independent structures can not be zero because   we have already set aside some of the independent couplings.   Since we have chosen the couplings in the structure $\Omega\Omega\Omega\Omega$ as independent couplings, we have to remove  all independent structures that are reproduced also by $\Omega\Omega\Omega\Omega$, \ie remove the terms that have four and more fields.  One finds the resulting  linear algebraic equations have no solution that involves only $\delta a_1,\cdots, \delta a_{5}$. It means there is no independent couplings at zero gauge field except the 4 couplings in  $\Omega\Omega\Omega\Omega$, \ie
\begin{eqnarray}
  {S} &\supset & -\frac{(2\pi\alpha')^2}{96}{T_p}\int {{d^{p + 1}}\sigma \sqrt { - \det {{\tilde G}_{ab}}  } } \Big[
  {b_1}\Omega _a^{\,\,\,\,c\nu}{\Omega ^{ab\mu}}\Omega _{b\,\,\,\,\nu}^{\,\,\,\,d}{\Omega _{cd\mu}}+ {b_2}\,\Omega _{a\,\,\,\,\mu}^{\,\,\,\,c}{\Omega ^{ab\mu}}\Omega _b^{\,\,\,\,d\nu}{\Omega _{cd\nu}}    \nonumber\\
&&\qquad\qquad\qquad\qquad\qquad\qquad + {b_3}\Omega _{ab}^{\,\,\,\,\,\,\nu}{\Omega ^{ab\mu}}{\Omega _{cd\nu}}\Omega _{\,\,\,\,\,\,\mu}^{cd} + {b_4}{\Omega _{ab\mu}}{\Omega ^{ab\mu}}{\Omega _{cd\nu}}{\Omega ^{cd\nu}}   \Big] \labell{aa11}
\end{eqnarray}
where we have  chosen the coefficients of the 4 independent couplings to be  $b_1,\cdots, b_4$.

Next, we consider  the couplings at  the level of two gauge fields at order $\alpha'^2$.  There are 18 independent couplings in the structure  $\Omega\Omega\Omega\Omega F F$ modulo the trace of $\Omega$. Apart from this structure, the Lagrangian in \reef{s1} has 4 structures as
\beqa
\cL' &\sim &DDFDDF+ DFDF\Omega\Omega+ FDF \Omega D\Omega+ FF D\Omega D\Omega
\labell{Action2}
\eeqa
Using the package xAct, one finds there are 118 gauge invariant couplings in these  structures. The vector in the total derivative \reef{tot} has 4 structures as
 \beqa
 \cI &\sim &FF\Omega D\Omega+FDF\Omega \Omega+DFDDF+FDDDF
\labell{TOT1}
\eeqa
The field redefinitions $\delta A_a$ and $\delta X^\mu$ in \reef{fred}  have 3 and 4 structures, respectively,  as
\beqa
\delta {A_a} &\sim&   \Omega \Omega DF + F\Omega D\Omega+DDDF \nonumber\\
\delta {X^\mu } &\sim&  \Omega DFDF + FDFD\Omega +F\Omega DDF+FFDD\Omega
  \labell{b60}
\eeqa
Using the package xAct, one can construct all possible contractions  in \reef{TOT1} and \reef{b60}. Then replacing them in \reef{SJK}, going to the local frame to write the equation \reef{SJK} in terms of the independent structures, and removing the the terms that have six and more fields which are reproduce also by the independent couplings in the structure  $\Omega\Omega\Omega\Omega F F$, one finds the resulting  linear algebraic equations has 4 solutions that involve only $\delta a_1,\cdots, \delta a_{118}$. It means there are 4 independent couplings at four gauge field level on top of the 18 couplings in the structure $\Omega\Omega\Omega\Omega F F$. One can set all of the coefficients  in \reef{s1} to zero except 4 of them. However, one is not totally free to choose the 4 couplings. The correct choices must be such that when one replaces the non-zero couplings in \reef{SJK}, the linear algebraic equations produces 4 relations $\delta a_1=\delta a_2=\delta a_3=\delta a_{4}=0$. For the wrong choices of the independent couplings, the algebraic equations, would produces less than 4 relations between only $\delta a_i$. There are different ways (schemes) to choose the 4 independent couplings.  One can choose the 4 independent couplings in the structure $DFDF\Omega\Omega$.  The couplings in a particular scheme are  the following:
\begin{eqnarray}
  {S} &\supset & -\frac{(2\pi\alpha')^2}{96}{T_p}\int {{d^{p + 1}}\sigma \sqrt { - \det {{\tilde G}_{ab}}  } } \Big[
  {a_1}{D_a}{F_{bc}}{D^a}{F^{bc}}{\Omega _{de\mu}}{\Omega ^{de\mu}} + {a_2}{D^a}{F^{bc}}{D^d}F_b^{\,\,e}\Omega _{ae}^{\,\,\,\,\,\,\mu}{\Omega _{cd\mu}}  \nonumber\\
  && + {a_3}{D^a}{F^{bc}}{D_b}F_a^{\,\,d}\Omega _c^{\,\,\,e\mu}{\Omega _{de\mu}} + {a_4}{D^a}{F^{bc}}{D^d}F_b^{\,\,e}\Omega _{ac}^{\,\,\,\,\,\,\mu}{\Omega _{de\mu}} \hfill \nonumber\\&&+ {f_1}\Omega _a^{\,\,\,\,e\mu}\Omega _b^{\,\,\,\,f\nu}{\Omega _{ce\nu}}{\Omega _{df\mu}}{F^{ab}}{F^{cd}}
   + {f_2}\Omega _a^{\,\,\,\,e\mu}\Omega _b^{\,\,\,\,f\nu}{\Omega _{ce\mu}}{\Omega _{df\nu}}{F^{ab}}{F^{cd}}\hfill \nonumber\\&& + {f_3}\Omega _a^{\,\,\,\,e\mu}\Omega _{b\,\,\,\,\mu}^{\,\,\,\,f}\Omega _{ce}^{\,\,\,\,\,\,\nu}{\Omega _{df\nu}}{F^{ab}}{F^{cd}}
   + {f_4}\Omega _a^{\,\,\,\,e\mu}\Omega _{be}^{\,\,\,\,\,\,\nu}\Omega _{c\,\,\,\,\mu}^{\,\,\,\,f}{\Omega _{df\nu}}{F^{ab}}{F^{cd}} \hfill \nonumber\\&& + {f_5}\Omega _{ac}^{\,\,\,\,\,\,\mu}\Omega _b^{\,\,\,\,e\nu}\Omega _{d\,\,\,\,\nu}^{\,\,\,\,f}{\Omega _{ef\mu}}{F^{ab}}{F^{cd}}
   + {f_6}\Omega _b^{\,\,\,\,d\mu}\Omega _c^{\,\,\,\,e\nu}\Omega _{d\,\,\,\,\nu}^{\,\,\,\,f}{\Omega _{ef\mu}}F_a^{\,\,c}{F^{ab}} \hfill \nonumber\\&& + {f_7}\Omega _c^{\,\,\,\,e\nu}{\Omega ^{cd\mu}}\Omega _{d\,\,\,\,\nu}^{\,\,\,\,f}{\Omega _{ef\mu}}{F_{ab}}{F^{ab}}
   + {f_8}\Omega _b^{\,\,\,\,d\mu}\Omega _c^{\,\,\,\,e\nu}\Omega _{d\,\,\,\,\mu}^{\,\,\,\,f}{\Omega _{ef\nu}}F_a^{\,\,c}{F^{ab}} \hfill \nonumber\\&&+ {f_9}\Omega _{ac}^{\,\,\,\,\,\,\mu}\Omega _{b\,\,\,\,\mu}^{\,\,\,\,e}\Omega _d^{\,\,\,\,f\nu}{\Omega _{ef\nu}}{F^{ab}}{F^{cd}}
   + {f_{10}}\Omega _b^{\,\,\,\,d\mu}\Omega _{c\,\,\,\,\mu}^{\,\,\,\,e}\Omega _d^{\,\,\,\,f\nu}{\Omega _{ef\nu}}F_a^{\,\,c}{F^{ab}} \hfill \nonumber\\&& + {f_{11}}\Omega _{c\,\,\,\,\mu}^{\,\,\,\,e}{\Omega ^{cd\mu}}\Omega _d^{\,\,\,\,f\nu}{\Omega _{ef\nu}}{F_{ab}}{F^{ab}}
 +{f_{12}}{\Omega _{bc}}^\mu{\Omega _d}^{f\nu}{\Omega ^{de}}_\mu{\Omega _{ef\nu}} F_a^{\,\,c}{F^{ab}} \nonumber\\&& + {f_{13}}\Omega _{ac}^{\,\,\,\,\,\,\mu}\Omega _{bd}^{\,\,\,\,\,\,\nu}{\Omega _{ef\nu}}\Omega _{\,\,\,\,\,\,\mu}^{ef}{F^{ab}}{F^{cd}}
 + {f_{14}}\Omega _b^{\,\,\,\,d\mu}\Omega _{cd}^{\,\,\,\,\,\,\nu}{\Omega _{ef\nu}}\Omega _{\,\,\,\,\,\,\mu}^{ef}F_a^{\,\,c}{F^{ab}} \hfill
  \nonumber\\&& + {f_{15}}\Omega _{cd}^{\,\,\,\,\,\,\nu}{\Omega ^{cd\mu}}{\Omega _{ef\nu}}\Omega _{\,\,\,\,\,\,\mu}^{ef}{F_{ab}}{F^{ab}} + {f_{16}}\Omega _{ac}^{\,\,\,\,\,\,\mu}{\Omega _{bd\mu}}{\Omega _{ef\nu}}{\Omega ^{ef\nu}}{F^{ab}}{F^{cd}} \hfill  \nonumber\\&&
   + {f_{17}}\Omega _b^{\,\,\,\,d\mu}{\Omega _{cd\mu}}{\Omega _{ef\nu}}{\Omega ^{ef\nu}}F_a^{\,\,c}{F^{ab}}+ {f_{18}}{\Omega _{cd\mu}}{\Omega ^{cd\mu}}{\Omega _{ef\nu}}{\Omega ^{ef\nu}}{F_{ab}}{F^{ab}}  \Big] \labell{a11}
\end{eqnarray}
where we have  chosen the coefficients of the 4 independent couplings to be $a_1,\cdots, a_4$. We have also included in above couplings the 18 independent couplings in the structure $\Omega\Omega\Omega\Omega F F$ with coefficients $f_1,\cdots, f_{18}$. Note that the above independent couplings become zero when $X^\mu$ is constant which is consistent with the fact that there is no independent couplings of two gauge fields at order $\alpha'^2$.

We now consider  the couplings that have four gauge fields  at order $\alpha'^2$.  Apart from  the structure $\Omega\Omega\Omega\Omega F FFF$, the Lagrangian in \reef{s1} has 6 structures as
\beqa
\cL' &\sim & DFDFDFDF+FDFDFDDF+FFDDFDDF \nonumber\\ &&+FFDFDF\Omega\Omega + FFFDF \Omega D\Omega + FFFF D\Omega D\Omega
\labell{Action4}
\eeqa
There are 1124 gauge invariant couplings in these structures. The vector in the total derivative \reef{tot} has 5 structures as
\beqa
\cI &\sim & FDFDFDF+FFDFDDF \nonumber\\ &&+FFFDDDF+FFFF\Omega D\Omega+FFFDF\Omega \Omega
\labell{TOT3}
\eeqa
The field redefinitions $\delta A_a$ and $\delta X^\mu$ in \reef{fred} each has 5 structures as
\beqa
\delta {A_a} &\sim&  DFDFDF + FDFDDF + FFDDDF + \Omega \Omega FFDF + FFF\Omega D\Omega \nonumber\\
\delta {X^\mu } &\sim&  FF\Omega DFDF + FFFDFD\Omega  + \Omega FFFDDF + FFFFDD\Omega  + \Omega \Omega \Omega FFFF
  \labell{b60c}
\eeqa
In this case, after removing the eight  and more  fields  from the independent structures in the local frame,  one finds the  resulting  linear algebraic equations have 68 solutions that involve only $\delta a_1,\cdots, \delta a_{1124}$. It means there are 68 independent couplings at four gauge field on top of the independent couplings in $\Omega\Omega\Omega\Omega F FFF$. We find that there are at least 4 independent couplings in the structures  in the first line of \reef{Action4} and there are at most 64 independent couplings in the structures in the second line of \reef{Action4}.   Since there are 4 independent couplings for only gauge field at order $\alpha'^2$, we choose the 4 couplings in structure $DFDFDFDF$ and 64 couplings in the structures in the second line of \reef{Action4}. The couplings  in a particular scheme are the following:
\begin{eqnarray}
  {S}&\!\!\!\!\!\!\! \supset \!\!\!\!\!\!\!\! & -\frac{(2\pi\alpha')^2}{96}T_p\!\int d^{p + 1}\sigma \sqrt{-\det {\tilde G}_{ab}}   \Big[ {c_1}{D_a}{F_{bc}}{D^a}{F^{bc}}{D_d}{F_{ef}}{D^d}{F^{ef}} + {c_2}{D_a}{F^{de}}{D^a}{F^{bc}}{D_f}{F_{de}}{D^f}{F_{bc}}  \nonumber\\
 && + {c_3}{D_a}{F^{de}}{D^a}{F^{bc}}{D_f}{F_{ce}}{D^f}{F_{bd}} + {c_4}{D_a}F_b^{\,\,d}{D^a}{F^{bc}}{D_e}{F_{df}}{D^e}F_c^{\,\,f} \labell{ac11}
 \hfill\\&& +{d_1}{D^a}{F^{bc}}{D^d}F_b^{\,\,e}{F_{fh}}{F^{fh}}\Omega _{ad}^{\,\,\,\,\,\,\mu}{\Omega _{ce\mu}}
   + {d_2}{D^a}{F^{bc}}{D_a}{F^{de}}{F_{fh}}{F^{fh}}\Omega _{bd}^{\,\,\,\,\,\,\mu}{\Omega _{ce\mu}}\hfill \nonumber\\&& + {d_3}{D^a}{F^{bc}}{D^d}{F^{ef}}F_b^{\,\,h}{F_{eh}}\Omega _{ad}^{\,\,\,\,\,\,\mu}{\Omega _{cf\mu}}
   + {d_4}{D^a}{F^{bc}}{D^d}{F^{ef}}F_a^{\,\,h}{F_{dh}}\Omega _{be}^{\,\,\,\,\,\,\mu}{\Omega _{cf\mu}} \hfill \nonumber\\&&+ {d_5}{D^a}{F^{bc}}{D_b}{F^{de}}F_d^{\,\,f}F_f^{\,\,h}\Omega _{ae}^{\,\,\,\,\,\,\mu}{\Omega _{ch\mu}}
   + {d_6}{D^a}{F^{bc}}{D^d}F_b^{\,\,e}{F_{fh}}{F^{fh}}\Omega _{ac}^{\,\,\,\,\,\,\mu}{\Omega _{de\mu}}  \hfill \nonumber\\&&+ {d_7}{D^a}{F^{bc}}{D^d}F_b^{\,\,e}F_a^{\,\,f}F_f^{\,\,h}\Omega _{ch}^{\,\,\,\,\,\,\mu}{\Omega _{de\mu}}
   + {d_8}{D^a}{F^{bc}}{D^d}{F^{ef}}F_b^{\,\,h}{F_{eh}}\Omega _{ac}^{\,\,\,\,\,\,\mu}{\Omega _{df\mu}} \hfill \nonumber\\&& + {d_9}{D^a}{F^{bc}}{D^d}{F^{ef}}F_a^{\,\,h}{F_{be}}\Omega _{ch}^{\,\,\,\,\,\,\mu}{\Omega _{df\mu}}
   + {d_{10}}{D^a}{F^{bc}}{D^d}{F^{ef}}{F_{ae}}F_b^{\,\,h}\Omega _{ch}^{\,\,\,\,\,\,\mu}{\Omega _{df\mu}}\hfill \nonumber\\&& + {d_{11}}{D^a}{F^{bc}}{D^d}{F^{ef}}{F_{ab}}F_e^{\,\,h}\Omega _{ch}^{\,\,\,\,\,\,\mu}{\Omega _{df\mu}}
   + {d_{12}}{D^a}{F^{bc}}{D^d}{F^{ef}}F_a^{\,\,h}{F_{be}}\Omega _{cf}^{\,\,\,\,\,\,\mu}{\Omega _{dh\mu}}\hfill \nonumber\\&& + {d_{13}}{D^a}{F^{bc}}{D^d}{F^{ef}}{F_{ae}}F_b^{\,\,h}\Omega _{cf}^{\,\,\,\,\,\,\mu}{\Omega _{dh\mu}}
   + {d_{14}}{D^a}{F^{bc}}{D^d}{F^{ef}}{F_{ab}}F_e^{\,\,h}\Omega _{cf}^{\,\,\,\,\,\,\mu}{\Omega _{dh\mu}}\hfill \nonumber\\&& + {d_{15}}{D^a}{F^{bc}}{D_a}F_b^{\,\,d}F_e^{\,\,h}{F^{ef}}\Omega _{cf}^{\,\,\,\,\,\,\mu}{\Omega _{dh\mu}}
   + {d_{16}}{D^a}{F^{bc}}{D_b}F_a^{\,\,d}F_e^{\,\,h}{F^{ef}}\Omega _{cf}^{\,\,\,\,\,\,\mu}{\Omega _{dh\mu}}\hfill \nonumber\\&& + {d_{17}}{D^a}{F^{bc}}{D^d}{F^{ef}}{F_{ae}}{F_{bf}}\Omega _c^{\,\,\,\,h\mu}{\Omega _{dh\mu}}
   + {d_{18}}{D^a}{F^{bc}}{D^d}{F^{ef}}{F_{ab}}{F_{ef}}\Omega _c^{\,\,\,\,h\mu}{\Omega _{dh\mu}}\hfill \nonumber\\&& + {d_{19}}{D^a}{F^{bc}}{D_a}F_b^{\,\,d}{F_{ef}}{F^{ef}}\Omega _c^{\,\,\,\,h\mu}{\Omega _{dh\mu}}
   + {d_{20}}{D^a}{F^{bc}}{D_b}F_a^{\,\,d}{F_{ef}}{F^{ef}}\Omega _c^{\,\,\,\,h\mu}{\Omega _{dh\mu}} \hfill \nonumber\\&&+ {d_{21}}{D^a}{F^{bc}}{D_b}{F^{de}}F_a^{\,\,f}F_d^{\,\,h}\Omega _{ch}^{\,\,\,\,\,\,\mu}{\Omega _{ef\mu}}
   + {d_{22}}{D^a}{F^{bc}}{D^d}F_b^{\,\,e}F_a^{\,\,f}F_d^{\,\,h}\Omega _{ch}^{\,\,\,\,\,\,\mu}{\Omega _{ef\mu}} \hfill \nonumber\\&& + {d_{23}}{D^a}{F^{bc}}{D_a}{F^{de}}F_b^{\,\,f}F_d^{\,\,h}\Omega _{ch}^{\,\,\,\,\,\,\mu}{\Omega _{ef\mu}}
   + {d_{24}}{D^a}{F^{bc}}{D_b}{F^{de}}F_d^{\,\,f}F_f^{\,\,h}\Omega _{ac}^{\,\,\,\,\,\,\mu}{\Omega _{eh\mu}} \hfill \nonumber\\&&+ {d_{25}}{D^a}{F^{bc}}{D_b}{F^{de}}F_a^{\,\,f}F_f^{\,\,h}\Omega _{cd}^{\,\,\,\,\,\,\mu}{\Omega _{eh\mu}}
   + {d_{26}}{D^a}{F^{bc}}{D^d}F_b^{\,\,e}F_a^{\,\,f}F_f^{\,\,h}\Omega _{cd}^{\,\,\,\,\,\,\mu}{\Omega _{eh\mu}} \hfill \nonumber\\&& + {d_{27}}{D^a}{F^{bc}}{D_a}{F^{de}}F_b^{\,\,f}F_f^{\,\,h}\Omega _{cd}^{\,\,\,\,\,\,\mu}{\Omega _{eh\mu}}
   + {d_{28}}{D^a}{F^{bc}}{D_b}{F^{de}}F_a^{\,\,f}F_d^{\,\,h}\Omega _{cf}^{\,\,\,\,\,\,\mu}{\Omega _{eh\mu}}  \hfill \nonumber\\&&+ {d_{29}}{D^a}{F^{bc}}{D_a}{F^{de}}F_b^{\,\,f}F_d^{\,\,h}\Omega _{cf}^{\,\,\,\,\,\,\mu}{\Omega _{eh\mu}}
   + {d_{30}}{D^a}{F^{bc}}{D_b}{F^{de}}{F_{ad}}{F^{fh}}\Omega _{cf}^{\,\,\,\,\,\,\mu}{\Omega _{eh\mu}}\hfill \nonumber\\&& + {d_{31}}{D^a}{F^{bc}}{D_a}{F^{de}}{F_{bd}}{F^{fh}}\Omega _{cf}^{\,\,\,\,\,\,\mu}{\Omega _{eh\mu}}
   + {d_{32}}{D^a}{F^{bc}}{D_b}{F^{de}}F_a^{\,\,f}{F_{df}}\Omega _c^{\,\,\,\,h\mu}{\Omega _{eh\mu}}\hfill \nonumber\\&& + {d_{33}}{D^a}{F^{bc}}{D^d}F_b^{\,\,e}F_a^{\,\,f}{F_{df}}\Omega _c^{\,\,\,\,h\mu}{\Omega _{eh\mu}}
   + {d_{34}}{D^a}{F^{bc}}{D_a}{F^{de}}F_b^{\,\,f}{F_{df}}\Omega _c^{\,\,\,\,h\mu}{\Omega _{eh\mu}}\hfill \nonumber\\&& + {d_{35}}{D^a}{F^{bc}}{D^d}F_b^{\,\,e}F_a^{\,\,f}F_c^{\,\,h}\Omega _{df}^{\,\,\,\,\,\,\mu}{\Omega _{eh\mu}}
   + {d_{36}}{D^a}{F^{bc}}{D_a}{F^{de}}F_b^{\,\,f}F_c^{\,\,h}\Omega _{df}^{\,\,\,\,\,\,\mu}{\Omega _{eh\mu}}\hfill \nonumber\\&& + {d_{37}}{D^a}{F^{bc}}{D_a}{F^{de}}{F_{bc}}{F^{fh}}\Omega _{df}^{\,\,\,\,\,\,\mu}{\Omega _{eh\mu}}
   + {d_{38}}{D^a}{F^{bc}}{D_a}F_b^{\,\,d}F_c^{\,\,e}{F^{fh}}\Omega _{df}^{\,\,\,\,\,\,\mu}{\Omega _{eh\mu}} \hfill \nonumber\\&&+ {d_{39}}{D^a}{F^{bc}}{D_b}F_a^{\,\,d}F_c^{\,\,e}{F^{fh}}\Omega _{df}^{\,\,\,\,\,\,\mu}{\Omega _{eh\mu}}
   + {d_{40}}{D^a}{F^{bc}}{D_b}{F_{ac}}{F^{de}}{F^{fh}}\Omega _{df}^{\,\,\,\,\,\,\mu}{\Omega _{eh\mu}}  \hfill \nonumber\\&&+ {d_{41}}{D^a}{F^{bc}}{D^d}F_b^{\,\,e}F_a^{\,\,f}F_d^{\,\,h}\Omega _{ce}^{\,\,\,\,\,\,\mu}{\Omega _{fh\mu}}
   + {d_{42}}{D^a}{F^{bc}}{D^d}{F^{ef}}{F_{ad}}{F_{be}}\Omega _c^{\,\,\,\,h\mu}{\Omega _{fh\mu}} \hfill \nonumber\\&&+ {d_{43}}{D^a}{F^{bc}}{D^d}F_b^{\,\,e}F_a^{\,\,f}F_c^{\,\,h}\Omega _{de}^{\,\,\,\,\,\,\mu}{\Omega _{fh\mu}}
   + {d_{44}}{D^a}{F^{bc}}{D_b}F_a^{\,\,d}F_c^{\,\,e}F_e^{\,\,f}\Omega _d^{\,\,\,\,h\mu}{\Omega _{fh\mu}}  \nonumber\\&&+ {d_{45}}{D^a}{F^{bc}}{D^d}F_b^{\,\,e}{F_{ad}}F_c^{\,\,f}\Omega _e^{\,\,\,\,h\mu}{\Omega _{fh\mu}} \hfill
   + {d_{46}}{D^a}{F^{bc}}{D_a}{F^{de}}{F_{bd}}F_c^{\,\,f}\Omega _e^{\,\,\,\,h\mu}{\Omega _{fh\mu}}  \hfill \nonumber\\&&+ {d_{47}}{D^a}{F^{bc}}{D^d}F_b^{\,\,e}{F_{ac}}F_d^{\,\,f}\Omega _e^{\,\,\,\,h\mu}{\Omega _{fh\mu}}
   + {d_{48}}{D^a}{F^{bc}}{D_a}{F^{de}}{F_{bc}}F_d^{\,\,f}\Omega _e^{\,\,\,\,h\mu}{\Omega _{fh\mu}} \hfill\nonumber\\&&+ {d_{49}}{D^a}{F^{bc}}{D_b}{F_{ac}}F_d^{\,\,f}{F^{de}}\Omega _e^{\,\,\,\,h\mu}{\Omega _{fh\mu}}
   + {d_{50}}{D^a}{F^{bc}}{D^d}F_b^{\,\,e}{F_{ad}}{F_{ce}}{\Omega _{fh\mu}}{\Omega ^{fh\mu}} \hfill \nonumber\\&&+ {d_{51}}{D^a}{F^{bc}}{D_a}{F^{de}}{F_{bd}}{F_{ce}}{\Omega _{fh\mu}}{\Omega ^{fh\mu}}
   + {d_{52}}{D^a}{F^{bc}}{D_a}{F^{de}}{F_{bc}}{F_{de}}{\Omega _{fh\mu}}{\Omega ^{fh\mu}} \hfill \nonumber\\&&+ {d_{53}}{D^a}{F^{bc}}{D_a}F_b^{\,\,d}F_c^{\,\,e}{F_{de}}{\Omega _{fh\mu}}{\Omega ^{fh\mu}}
   + {d_{54}}{D^a}{F^{bc}}{D_b}F_a^{\,\,d}F_c^{\,\,e}{F_{de}}{\Omega _{fh\mu}}{\Omega ^{fh\mu}} \hfill \nonumber\\&& + {d_{55}}{D^a}{F^{bc}}{D_b}{F_{ac}}{F_{de}}{F^{de}}{\Omega _{fh\mu}}{\Omega ^{fh\mu}}
 + {d_{56}}{D^a}{F^{bc}}{D_a}F_b^{\,\,d}F_e^{\,\,h}{F^{ef}}\Omega _{cd}^{\,\,\,\,\,\,\,\mu}{\Omega _{fh\mu}} \nonumber\\&&+ {d_{57}}{D^a}{F^{bc}}{D_a}{F^{de}}F_b^{\,\,f}F_d^{\,\,h}\Omega _{ce}^{\,\,\,\,\,\,\,\mu}{\Omega _{fh\mu}}
 + {d_{58}}{D^a}{F^{bc}}{D^d}{F^{ef}}{F_{ab}}F_c^{\,\,h}\Omega _{de}^{\,\,\,\,\,\,\,\mu}{\Omega _{fh\mu}} \nonumber\\&& + {d_{59}}{D^a}{F^{bc}}{D^d}{F^{ef}}F_a^{\,\,h}{F_{bh}}\Omega _{ce}^{\,\,\,\,\,\,\,\mu}{\Omega _{df\mu}}
 + {d_{60}}{D^a}{F^{bc}}{D^d}F_b^{\,\,e}F_a^{\,\,f}F_d^{\,\,h}\Omega _{cf}^{\,\,\,\,\,\,\,\mu}{\Omega _{eh\mu}}\nonumber\\&& + {d_{61}}{D^a}{F^{bc}}{D^d}F_b^{\,\,e}{F_{ad}}{F^{fh}}\Omega _{cf}^{\,\,\,\,\,\,\mu}{\Omega _{eh\mu}}
 + {d_{62}}{D^a}{F^{bc}}{D^d}F_b^{\,\,e}F_a^{\,\,f}{F_{cf}}\Omega _d^{\,\,\,\,h\mu}{\Omega _{eh\mu}}\nonumber\\&& + {d_{63}}{D^a}{\Omega ^{bc\mu}}{D^d}\Omega _{\,\,\,\,\,\,\mu}^{ef}{F_{ad}}{F_{be}}F_c^{\,\,h}{F_{fh}}
 + {d_{64}}{D^a}{F^{bc}}{D^d}{\Omega ^{ef\mu}}{F_{ab}}F_e^{\,\,h}{F_{fh}}{\Omega _{cd\mu}}
+\Omega\Omega\Omega\Omega F FFF \Big] \nn
 \end{eqnarray}
where we have  chosen the coefficients of the 4 independent couplings to be $c_1,\cdots, c_4$, and the 64 couplings to be  $d_1,\cdots, d_{64}$.  Since the couplings in the structure $\Omega\Omega\Omega\Omega F FFF $ involve eight   fields $F,\Omega$  in which we are not interested in this paper, we did not write  the dependent couplings in this structure. Note that when $\Omega$ is zero, the above couplings reduces to the independent couplings of four gauge field at order $\alpha'^2$.

We finally consider in this section the couplings which have six gauge fields. Apart from  the structure $\Omega\Omega\Omega\Omega F FFF FF$, the Lagrangian in \reef{s1} has 6 structures as
\beqa
\cL' &\sim & FFDFDFDFDF+FFFDFDFDDF+FFFFDDFDDF \nonumber\\ &&+FFFFDFDF\Omega\Omega + FFFFFDF \Omega D\Omega + FFFFFF D\Omega D\Omega
\labell{Action42}
\eeqa
In this case the couplings in the structures in the second line have eight gauge field or the  second fundamental form in which we are not interested in this paper. On the other hand, in the scheme that we are using in this paper in which the independent couplings should be reduced to the independent couplings of only gauge field when $X^\mu$ is a constant, one can find the independent couplings in the first line by finding the independent couplings of  only the gauge field. At the end, the partial derivatives are replaced by the covariant derivatives.  So we consider only the six gauge field structures
\beqa
\cL_F' &\sim & FFDFDFDFDF+FFFDFDFDDF+FFFFDDFDDF \nonumber
\labell{Action43}
\eeqa
There are 2836 gauge invariant couplings in these structures. The vector in the total derivative \reef{tot} has 3 structures as
\beqa
\cI &\sim & FFFDFDFDF+FFFFDFDDF+FFFFFDDDF
\labell{TOT31}
\eeqa
The field redefinition $\delta A_a$  has 3 structures as
\beqa
\delta {A_a} &\sim&  FFDFDFDF + FFFDFDDF + FFFFDDDF
  \labell{b60c1}
\eeqa
The derivatives are all partial derivatives. In this case, one needs only to impose the Bianchi identity \reef{bian} to find the corresponding  independent structures in \reef{SJK}. One finds the   linear algebraic equations have 64 solutions that involve only $\delta a_1,\cdots, \delta a_{2836}$. It means there are 64 independent couplings at six gauge fields when $X^\mu$ is constant.  The couplings  in a particular scheme are the following:
\begin{eqnarray}
  {S} &\supset &   -\frac{(2\pi\alpha')^2}{96}{T_p}\int {{d^{p + 1}}\sigma \sqrt { - \det {{\tilde G}_{ab}}  } } \Big[
  {e_1}{D^a}{F^{bc}}{D_b}{F^{de}}{D^f}{F_{de}}{D^h}F_f^{\,\,u}{F_{au}}{F_{ch}} \hfill \nonumber\\&&
   + {e_2}{D^a}{F^{bc}}{D_b}{F^{de}}{D_d}{F^{fh}}{D^u}{F_{ef}}{F_{au}}{F_{ch}} + {e_3}{D^a}{F^{bc}}{D_b}{F^{de}}{D_d}{F^{fh}}{D^u}{F_{fh}}{F_{ae}}{F_{cu}} \hfill \nonumber\\&&
   + {e_4}{D^a}{F^{bc}}{D_b}F_c^{\,\,d}{D^e}{F^{fh}}{D^u}{F_{fh}}{F_{ae}}{F_{du}} + {e_5}{D^a}{F^{bc}}{D_b}F_a^{\,\,d}{D^e}{F^{fh}}{D_f}F_h^{\,\,u}{F_{ce}}{F_{du}} \hfill \nonumber\\&&
   + {e_6}{D^a}{F^{bc}}{D_a}F_b^{\,\,d}{D^e}{F^{fh}}{D_f}{F_{eh}}F_c^{\,\,u}{F_{du}} + {e_7}{D^a}{F^{bc}}{D_b}{F^{de}}{D^f}{F_{cd}}{D^h}F_f^{\,\,u}{F_{au}}{F_{eh}} \hfill \nonumber\\&&
   + {e_8}{D^a}{F^{bc}}{D_b}{F^{de}}{D_d}{F^{fh}}{D^u}{F_{cf}}{F_{au}}{F_{eh}} + {e_9}{D^a}{F^{bc}}{D_a}{F^{de}}{D_f}F_d^{\,\,u}{D^f}F_b^{\,\,h}{F_{cu}}{F_{eh}} \hfill \nonumber\\&&
   + {e_{10}}{D^a}{F^{bc}}{D_a}{F^{de}}{D_b}F_d^{\,\,f}{D^h}F_f^{\,\,u}{F_{cu}}{F_{eh}} + {e_{11}}{D^a}{F^{bc}}{D_b}F_a^{\,\,d}{D^e}F_c^{\,\,f}{D^h}F_f^{\,\,u}{F_{du}}{F_{eh}} \hfill \nonumber\\&&
   + {e_{12}}{D^a}{F^{bc}}{D_b}{F^{de}}{D_d}{F^{fh}}{D^u}{F_{fh}}{F_{ac}}{F_{eu}} + {e_{13}}{D^a}{F^{bc}}{D_b}{F^{de}}{D_c}{F^{fh}}{D_d}F_f^{\,\,u}{F_{ah}}{F_{eu}} \hfill \nonumber\\&&
   + {e_{14}}{D^a}{F^{bc}}{D_b}{F^{de}}{D_d}{F^{fh}}{D_f}F_c^{\,\,u}{F_{ah}}{F_{eu}} + {e_{15}}{D^a}{F^{bc}}{D_b}{F^{de}}{D_d}F_c^{\,\,f}{D_f}{F^{hu}}{F_{ah}}{F_{eu}} \hfill \nonumber\\&&
   + {e_{16}}{D^a}{F^{bc}}{D_b}F_c^{\,\,d}{D^e}F_d^{\,\,f}{D_f}{F^{hu}}{F_{ah}}{F_{eu}} + {e_{17}}{D^a}{F^{bc}}{D_b}{F^{de}}{D_d}F_c^{\,\,f}{D^h}F_f^{\,\,u}{F_{ah}}{F_{eu}} \hfill \nonumber\\&&
   + {e_{18}}{D^a}{F^{bc}}{D_b}F_c^{\,\,d}{D^e}F_d^{\,\,f}{D^h}F_f^{\,\,u}{F_{ah}}{F_{eu}} + {e_{19}}{D^a}{F^{bc}}{D_b}{F^{de}}{D^f}{F_{cd}}{D^h}F_f^{\,\,u}{F_{ah}}{F_{eu}} \hfill \nonumber\\&&
   + {e_{20}}{D^a}{F^{bc}}{D_b}{F^{de}}{D_d}{F^{fh}}{D^u}{F_{cf}}{F_{ah}}{F_{eu}} + {e_{21}}{D^a}{F^{bc}}{D_b}{F^{de}}{D^f}F_c^{\,\,h}{D^u}{F_{df}}{F_{ah}}{F_{eu}} \hfill \nonumber\\&&
   + {e_{22}}{D^a}{F^{bc}}{D_b}{F^{de}}{D_c}{F^{fh}}{D_d}{F_{fh}}F_a^{\,\,u}{F_{eu}} + {e_{23}}{D^a}{F^{bc}}{D_b}F_a^{\,\,d}{D^e}F_c^{\,\,f}{D^h}F_d^{\,\,u}{F_{eh}}{F_{fu}} \hfill \nonumber\\&&
   + {e_{24}}{D^a}{F^{bc}}{D_a}{F^{de}}{D_b}{F^{fh}}{D^u}{F_{de}}{F_{cf}}{F_{hu}} + {e_{25}}{D^a}{F^{bc}}{D_b}F_a^{\,\,d}{D^e}F_c^{\,\,f}{D^h}F_e^{\,\,u}{F_{df}}{F_{hu}} \hfill \nonumber\\&&
   + {e_{26}}{D^a}{F^{bc}}{D_b}F_a^{\,\,d}{D^e}F_c^{\,\,f}{D_f}F_e^{\,\,h}F_d^{\,\,u}{F_{hu}} + {e_{27}}{D^a}{F^{bc}}{D_b}F_a^{\,\,d}{D^e}F_c^{\,\,f}{D^h}{F_{ef}}F_d^{\,\,u}{F_{hu}} \hfill \nonumber\\&&
   + {e_{28}}{D^a}{F^{bc}}{D_b}{F^{de}}{D_c}{F^{fh}}{D_d}F_a^{\,\,u}{F_{ef}}{F_{hu}} + {e_{29}}{D^a}{F^{bc}}{D_a}{F_{bc}}{D_d}{F^{hu}}{D^d}{F^{ef}}{F_{ef}}{F_{hu}} \hfill \nonumber\\&&
   + {e_{30}}{D^a}{F^{bc}}{D_b}F_a^{\,\,d}{D_c}F_d^{\,\,e}{D^f}{F^{hu}}{F_{ef}}{F_{hu}} + {e_{31}}{D^a}{F^{bc}}{D_b}F_c^{\,\,d}{D_d}F_a^{\,\,e}{D^f}{F^{hu}}{F_{ef}}{F_{hu}} \hfill \nonumber\\&&
   + {e_{32}}{D^a}{F^{bc}}{D_a}{F^{de}}{D_b}F_d^{\,\,f}{D^h}F_c^{\,\,u}{F_{ef}}{F_{hu}} + {e_{33}}{D^a}{F^{bc}}{D_b}{F^{de}}{D_d}F_a^{\,\,f}{D^h}F_c^{\,\,u}{F_{ef}}{F_{hu}} \hfill \nonumber\\&&
   + {e_{34}}{D^a}{F^{bc}}{D_a}{F^{de}}{D^f}{F_{bd}}{D^h}F_c^{\,\,u}{F_{ef}}{F_{hu}} + {e_{35}}{D^a}{F^{bc}}{D_a}F_b^{\,\,d}{D^e}F_c^{\,\,f}{D^h}F_d^{\,\,u}{F_{ef}}{F_{hu}} \hfill \nonumber\\&&
   + {e_{36}}{D^a}{F^{bc}}{D_b}F_a^{\,\,d}{D^e}F_c^{\,\,f}{D^h}F_d^{\,\,u}{F_{ef}}{F_{hu}} + {e_{37}}{D^a}{F^{bc}}{D_a}{F^{de}}{D^f}{F_{bc}}{D^h}F_d^{\,\,u}{F_{ef}}{F_{hu}} \hfill \nonumber\\&&
   + {e_{38}}{D^a}{F^{bc}}{D_a}{F^{de}}{D_f}F_c^{\,\,h}{D^f}{F_{bd}}F_e^{\,\,u}{F_{hu}} + {e_{39}}{D^a}{F^{bc}}{D_a}{F^{de}}{D^f}{F_{bd}}{D^h}{F_{cf}}F_e^{\,\,u}{F_{hu}} \hfill \nonumber\\&&
   + {e_{40}}{D^a}{F^{bc}}{D_a}{F^{de}}{D_b}F_d^{\,\,f}{D_e}F_c^{\,\,h}F_f^{\,\,u}{F_{hu}} + {e_{41}}{D^a}{F^{bc}}{D_a}F_b^{\,\,d}{D_c}{F^{ef}}{D_e}F_d^{\,\,h}F_f^{\,\,u}{F_{hu}} \hfill \nonumber\\&&
   + {e_{42}}{D^a}{F^{bc}}{D_b}F_a^{\,\,d}{D_c}{F^{ef}}{D_e}F_d^{\,\,h}F_f^{\,\,u}{F_{hu}} + {e_{43}}{D^a}{F^{bc}}{D_b}{F_{ac}}{D^d}{F^{ef}}{D_e}F_d^{\,\,h}F_f^{\,\,u}{F_{hu}} \hfill \nonumber\\&&
   + {e_{44}}{D^a}{F^{bc}}{D_a}F_b^{\,\,d}{D_e}F_d^{\,\,h}{D^e}F_c^{\,\,f}F_f^{\,\,u}{F_{hu}} + {e_{45}}{D^a}{F^{bc}}{D_b}{F^{de}}{D_d}F_a^{\,\,f}{D^h}{F_{ce}}F_f^{\,\,u}{F_{hu}} \hfill \nonumber\\&&
   + {e_{46}}{D^a}{F^{bc}}{D_a}{F^{de}}{D^f}{F_{bd}}{D^h}{F_{ce}}F_f^{\,\,u}{F_{hu}} + {e_{47}}{D^a}{F^{bc}}{D_a}F_b^{\,\,d}{D_c}{F^{ef}}{D^h}{F_{de}}F_f^{\,\,u}{F_{hu}} \hfill \nonumber\\&&
   + {e_{48}}{D^a}{F^{bc}}{D_b}F_a^{\,\,d}{D_c}{F^{ef}}{D^h}{F_{de}}F_f^{\,\,u}{F_{hu}} + {e_{49}}{D^a}{F^{bc}}{D_a}{F^{de}}{D^f}{F_{bc}}{D^h}{F_{de}}F_f^{\,\,u}{F_{hu}} \hfill \nonumber\\&&
   + {e_{50}}{D^a}{F^{bc}}{D_b}{F_{ac}}{D^d}{F^{ef}}{D_e}{F_{df}}{F_{hu}}{F^{hu}} + {e_{51}}{D^a}{F^{bc}}{D_a}F_b^{\,\,d}{D^e}F_c^{\,\,f}{D_f}{F_{de}}{F_{hu}}{F^{hu}} \hfill \nonumber\\&&
   + {e_{52}}{D^a}{F^{bc}}{D_b}F_a^{\,\,d}{D^e}F_c^{\,\,f}{D_f}{F_{de}}{F_{hu}}{F^{hu}} + {e_{53}}{D^a}{F^{bc}}{D_a}{F^{de}}{D_f}{F_{de}}{D^f}{F_{bc}}{F_{hu}}{F^{hu}} \hfill \nonumber\\&&
   + {e_{54}}{D^a}{F^{bc}}{D_a}{F^{de}}{D_f}{F_{ce}}{D^f}{F_{bd}}{F_{hu}}{F^{hu}} +
{e_{55}}{D^a}{F^{bc}}{D_a}{F^{de}}{D_f}F_d^{\,\,u}{D^f}F_b^{\,\,h}{F_{ce}}{F_{hu}} \nonumber\\&&+ {e_{56}}{D_e}{D^h}F_c^{\,\,u}{D_f}{D_u}{F_{dh}}F_a^{\,\,c}{F^{ab}}F_b^{\,\,d}{F^{ef}} + {e_{57}}{D_a}{D_c}{F_{eh}}{D_b}{D_f}{F_{du}}{F^{ab}}{F^{cd}}{F^{ef}}{F^{hu}} \nonumber\\&&
 + {e_{58}}{D^a}{F^{bc}}{D_a}{F^{de}}{D_b}{F^{fh}}{D^u}{F_{df}}{F_{ce}}{F_{hu}} + {e_{59}}{D^a}{F^{bc}}{D_b}F_a^{\,\,d}{D^e}{F^{fh}}{D_f}F_e^{\,\,u}{F_{ch}}{F_{du}} \nonumber\\&&
 + {e_{60}}{D^a}{F^{bc}}{D_b}F_a^{\,\,d}{D_c}F_d^{\,\,e}{D^f}F_e^{\,\,h}F_f^{\,\,u}{F_{hu}} + {e_{61}}{D^a}{F^{bc}}{D_b}F_c^{\,\,d}{D_d}F_a^{\,\,e}{D^f}F_e^{\,\,h}F_f^{\,\,u}{F_{hu}} \nonumber\\&&
 + {e_{62}}{D^a}{F^{bc}}{D_b}{F^{de}}{D^f}F_c^{\,\,h}{D^u}{F_{dh}}{F_{af}}{F_{eu}} + {e_{63}}{D_b}{D^h}F_e^{\,\,u}{D_f}{D_u}{F_{ch}}F_a^{\,\,c}{F^{ab}}F_d^{\,\,f}{F_{de}} \nonumber\\&&
+ {e_{64}}{D_e}{D_h}{F_{fu}}{D^a}{F^{bc}}{D_b}{F_{ac}}F_d^{\,\,f}{F^{de}}{F^{hu}}+\Omega\Omega\Omega\Omega F FFF FF \nn\\&&
+FFFFDFDF\Omega\Omega + FFFFFDF \Omega D\Omega + FFFFFF D\Omega D\Omega \Big]  \labell{ab11}
\end{eqnarray}
where $e_1,\cdots, e_{64}$ are some  parameters. The couplings  in the structures in the last line above  and in the structure $\Omega\Omega\Omega\Omega F FFF FF $ involve more than six  gauge field and/or the second fundamental form in which we are not interested in this paper. When $\Omega$ is zero, the above couplings reduces to the independent couplings of four gauge field at order $\alpha'^2$. Hence, the derivatives in the above independent couplings are now covariant derivatives.

The parameters of  the independent couplings in \reef{aa11}, \reef{a11}, \reef{ac11} and \reef{ab11} are  background independent  parameters which may be  found by the appropriate  S-matrix elements in flat spacetime. The couplings of four gauge field and/or the second fundamental form have been found by the S-matrix element of four open string vertex operators  \cite{Andreev:1988cb, Garousi:2015qgr}. They are
\beqa
&&a_1=\frac{1}{2}\,\,,\,\,a_2=a_3=a_4=-2;\,\,b_1=0\,\,,\,\,b_2=2\,\,,\,\,b_3=-2\,\,,\,\,b_4=0;
\nonumber\\
&&c_1=\frac{1}{8}\,\,,\,\,c_2=\frac{1}{4}\,\,,\,\,c_3=-\frac{1}{2}\,\,,\,\,c_4=-1
\labell{b4}
\eeqa
However, we are going to find the parameters  in the next section  by imposing the T-duality constraint.

\section{T-duality constraint}

We now try to fix the parameters in the actions \reef{aa11}, \reef{a11}, \reef{ac11} and \reef{ab11}. The assumption that the world volume effective action at the critical dimension is background independent, means the parameters in these actions are independent of the background. Hence, to fix them we consider a specific background which has a circle. That is, the   manifold has the structure $M^{(10)} = M^{(9)} \times S^{(1)}$.
 The manifold $M^{(10)}$ has coordinates $x^\mu = (x^{\tilde{\mu}}, y)$ where $x^{\tilde{\mu}}$ is the coordinates of the manifold $M^{(9)}$, and $y$ is the coordinate of the circle $S^{(1)}$. The world volume action has two reductions on the circle. When the $D_p$-brane is along the circle, \ie $a=(\tilde{a},y)$, the reduction is called $S_p^w$, and when the $D_p$-brane is orthogonal to the circle,  \ie $a=\tilde{a}$, the
 reduction is called $S_p^t$.  These two actions are not identical. However, the transformation   of $S_p^w$ under the following T-duality transformations
\beqa
{A_y} &\rightarrow & {X ^y},\nonumber\\
{A_{\tilde a}} &\rightarrow & {A_{\tilde a}},\nonumber\\
{X ^{\tilde \mu }}&\rightarrow &{X ^{\tilde \mu }}
\labell{a20}
\eeqa
 which is called $S_{p-1}^{wT}$, should be the same as $S_{p-1}^t$, up to some total derivative terms and field redefinitions in the base space, \ie
 \beqa
\Delta \tilde{S}+\tilde{\cJ}+\tilde{\cK}&=&0\labell{tSJK}
\eeqa
where  $\Delta \tilde{S} = S_{p-1}^{wT}-S_{p-1}^t $, the total derivative term
 $\tilde{\cJ}$ and the field redefinition contributions are
\beqa
\tilde{\cJ}&=&\alpha'^2 T_{p-1}\int d^{p}\sigma\, \sqrt { - \det \tilde{g}_{\ta\tb}} \tilde D_{\tilde{a}}\tilde{\cI}^{\tilde{a}}\labell{ttot}\\
\tilde{\cK}&=&\alpha'^2T_{p-1}\int d^{p}\sigma\,
 \sqrt { - \det \tilde{g}_{\ta\tb}}  \Bigg[
- {{\tilde D}_{\tilde a}}{F^{\tilde a\tilde b}}\delta {A_{\tilde b}} - {{\tilde g}^{\tilde a\tilde b}}{\tilde{\Omega} _{\tilde a\tilde b}}^{\tilde \nu }\delta {X^{\tilde \mu }}{\eta _{\tilde \mu \tilde \nu }} - {{\tilde g}^{\tilde a\tilde b}}{\tilde{\Omega} _{\tilde a\tilde b}}^y\delta {X^y}{\eta _{yy}} +  \cdots  \Bigg ]\nn
\eeqa
where $\tilde{\cI}^{\tilde{a}}$ is a  vector which is made of the base space fields $F,\tilde{\Omega}^{\tilde{\mu}},\prt X^y$ and their covariant derivatives at order $\alpha'^{3/2}$ with coefficients $j_1,j_2,\cdots$. In above equations, the world volume indices are contracted with the inverse of the pull-back of the base space metric onto the world volume of $D_{p-1}$-brane, \ie
\beqa
\tilde{g}_{\tilde{a}\tilde{b}}&=& \prt_{\tilde{a}} X^{\tilde{\mu}}\prt_{\tilde{b}} X^{\tilde{\nu}} \eta_{\tilde{\mu}\tilde{\nu}}\labell{tpull}
\eeqa
and dots in $\tilde{\cK}$ represent the terms which involve all higher orders of $F$ and $\prt X^y$ which are resulted from inserting in the world volume reduction of \reef{a1}, the following field redefinitions:
\beqa
A_{\tilde{a}} &\rightarrow & A_{\tilde{a}}+\alpha'^{3/2}\delta A_{\tilde{a}}, \nonumber\\
X^{\tilde{\mu }} &\rightarrow & X^{\tilde{\mu } } + \alpha'^{3/2}\delta X^{\tilde{\mu }} \nonumber\\
X^y &\rightarrow & X^y + \alpha'^{3/2}\delta X^y
\labell{tb40}
\eeqa
and using  integration by part. The coefficients of the gauge invariant terms in $\delta A_{\tilde{a}}$, $\delta X^{\tilde{\mu }}$, $\delta X^y$  at order $\alpha'^{3/2}$ are $k_1,k_2,\cdots$. Unlike in $\cK$, the dots in $\tilde{\cK}$ can not be ignored because they have contribution with some  fixed parameters in some of  the structures in the constraint \reef{tSJK}, \ie if one ignores them, then one would find the world volume actions   \reef{aa11}, \reef{a11}, \reef{ac11} and \reef{ab11} satisfy the constraint  \reef{tSJK} when all  parameters in the actions are zero which is not true.

For the world-volume reduction, $a=(\tilde{a},y)$ and  $\mu=(\tilde{\mu},y)$. Using the fact that the second fundamental form is zero when $\mu$ is a world volume index,  and the fact that in the dimensional reduction one assumes field are independent of the $y$ coordinate, \ie the Kaluza-Klein modes are ignored,  one finds the  following non-zero world volume reductions:
\beqa
{{\tilde G}_{\ta\tb}} &=& {{\tilde g}_{\tilde a\tilde b}}\nonumber\\
{\Omega _{\ta\tb}}^{\tilde\mu}  &=& {{\tilde \Omega }_{\tilde a\tilde b}}^{\ \ \tilde \mu }\nonumber\\
{D_\ta}{\Omega _{\tb\tc}}^{\tilde\mu}  &=& {{\tilde D}_{\tilde a}}{{\tilde \Omega }_{\tilde b\tilde c}}^{\ \ \tilde \mu }\nonumber\\
{F_{\ta\tb}} &=& {F_{\tilde a\tilde b}}\nonumber\\
{F_{\ta y}} &=& {F_{\tilde ay}}\nonumber\\
{F_{y\ta}} &=& {F_{y\tilde a}}\nonumber\\
{D_\ta}{F_{\tb\tc}} &=& {{\tilde D}_{\tilde a}}{F_{\tilde b\tilde c}}\nonumber\\
{D_\ta}{F_{\tb y}} &=& {{\tilde D}_{\tilde a}}{F_{\tilde by}}\nonumber\\
{D_\ta}{F_{y\tb}} &=& {{\tilde D}_{\tilde a}}{F_{y\tilde b}}\nonumber\\
{D_\ta}{D_\tb}{F_{\tc\td}} &=& {{\tilde D}_{\tilde a}}{{\tilde D}_{\tilde b}}{F_{\tilde c\tilde d}}\nonumber\\
{D_\ta}{D_\tb}{F_{\tc y}} &=& {{\tilde D}_{\tilde a}}{{\tilde D}_{\tilde b}}F_{\tc y}\nonumber\\
{D_\ta}{D_\tb}{F_{y\tc}} &=&   {{\tilde D}_{\tilde a}}{{\tilde D}_{\tilde b}}F_{y\tc}
\labell{a21}
\eeqa
where ${{\tilde \Omega }_{\tilde a\tilde b}}^{\ \ \tilde \mu } = {{\tilde D}_{\tilde a}}{\partial _{\tilde b}}{X ^{\tilde \mu }}$, and the covariant derivatives on the right-hand side are made of the pull-back metric \reef{tpull}.

For the transverse  reduction, $a=\tilde{a}$ and  $\mu=(\tilde{\mu},y)$. Since the index $y$ is a transverse index, one finds the following  non-zero transverse reductions:
\beqa
{{\tilde G}_{\ta\tb}} &\to& {{\tilde g}_{\tilde a\tilde b}} + {\partial _{\tilde a}}{X ^y}{\partial _{\tilde b}}{X ^y}\nonumber\\
{\Omega _{\ta\tb}}^{\tilde\mu}  &\to& {{\tilde \Omega }_{\tilde a\tilde b}}^{\ \ \tilde \mu } - {\partial _{\tilde c}}{X ^y}{\partial ^{\tilde c}}{X ^{\tilde \mu }}{{\tilde \Omega }_{\tilde a\tilde b}}^{\ \ y}(\frac{1}{{1 - {\partial _{\tilde e}}{X ^y}{\partial ^{\tilde e}}{X ^y}}})\nonumber\\
{\Omega _{\ta\tb}}^y &\to& {{\tilde \Omega }_{\tilde a\tilde b}}^{\ \ y}(\frac{1}{{1 - {\partial _{\tilde e}}{X ^y}{\partial ^{\tilde e}}{X ^y}}})\nonumber\\
{D_\ta}{\Omega _{\tb\tc}}^{\tilde\mu}  &\to& {{\tilde D}_{\tilde a}}{{ \Omega }_{\tilde b\tilde c}}{}^{ \tilde \mu } - {\partial ^{\tilde d}}{X ^y}({{ \Omega }_{\tilde c\tilde d}}{}^{ \tilde \mu }{{ \Omega }_{\tilde a\tilde b}}{}^{ y} + {{ \Omega }_{\tilde b\tilde d}}^{ \tilde \mu }{{ \Omega }_{\tilde a\tilde c}}{}^{ y})\nonumber\\
{D_\ta}{\Omega _{\tb\tc}}^y &\to& {{\tilde D}_{\tilde a}}{{ \Omega }_{\tilde b\tilde c}}^{y} - {\partial ^{\tilde d}}{X ^y}({{ \Omega }_{\tilde c\tilde d}}^{ y}{{ \Omega }_{\tilde a\tilde b}}^{ y} + {{ \Omega }_{\tilde b\tilde d}}^{ y}{{ \Omega }_{\tilde a\tilde c}}^{ y})\nonumber\\
{F_{\ta\tb}} &\to& {F_{\tilde a\tilde b}}\nonumber\\
{D_\ta}{F_{\tb\tc}} &\to& {{\tilde D}_{\tilde a}}{F_{\tilde b\tilde c}} + {\partial ^{\tilde d}}{X ^y}({F_{\tilde c\tilde d}}{{\Omega }_{\tilde a\tilde b}}^{ y} - {F_{\tilde b\tilde d}}{{ \Omega }_{\tilde a\tilde c}}^{ y})\nonumber\\
{D_\ta}{D_\tb}{F_{\tc\td}} &\to& {{\tilde D}_{\tilde a}}{{ D}_{\tilde b}}{F_{\tilde c\tilde d}} - {\partial ^{\tilde e}}{X ^y}({{ D}_{\tilde e}}{F_{\tilde c\tilde d}}{{ \Omega }_{\tilde a\tilde b}}^{ y} - {{ D}_{\tilde b}}{F_{\tilde d\tilde e}}{{ \Omega }_{\tilde a\tilde c}}^{ y} + {{ D}_{\tilde b}}{F_{\tilde c\tilde e}}{{ \Omega }_{\tilde a\tilde d}}^{ y})
\labell{a22}
\eeqa
where, for the simplicity in writing,  on the right-hand side of the transverse reductions of $D\Omega$ and $DDF$, we have not written the results completely in terms of the base space tensors  $\tilde \Omega$ and $\tilde D$.  One can easily replace them from the reductions of $\Omega$ and $DF$. In both world volume and transverse reductions, one observers that the identity \reef{Identity} reduces to the corresponding identity in the base space, \ie
 \beqa
\tilde{\Omega }_{\tilde a\tilde b}{}^{\tilde \mu }{\partial _{\tilde c}}{X^{\tilde \nu }}\eta_{\tilde \mu\tilde \nu} = 0 \labell{Identityred}
\eeqa
and the reductions satisfies the Bianchi identity \reef{bian}. Note that the gauge field in the base space satisfies its corresponding Bianchi identity
\beqa
\prt_{[\ta}F_{\tb\tc]}=0\labell{tbian}
\eeqa
Note also that there is no relation corresponding to  \reef{Identity} for $\mu,\nu=y$. Hence,  one can not remove the term $\prt_\ta X^y$ from the independent covariant couplings in the base space.

Using the  reductions \reef{a21} and \reef{a22}, one can calculate $\Delta\tilde S$ in \reef{tSJK}. To solve the T-duality constraint \reef{tSJK}, one has to write it in terms of independent couplings in the base space, \ie the Bianchi identity \reef{tbian} and the identities corresponding to the second fundamental forms must be imposed into it. As in the previous section,  we write the covariant derivatives in the base  space in terms of partial derivatives and   the  Levi-Civita connection which is made of the pull-back metric \reef{tpull}. Moreover, one can go to the local frame in which the  Levi-Civita connection is zero but its partial derivatives are not zero. Then, one can  write the derivatives of the connection in terms of the pull-back metric \reef{tpull}. In the resulting expression,  then one has to replace  the two $\prt_\ta X^{\tilde \mu}$ in which their spacetime index are contracted with each other, \ie $\prt_\ta X^{\tilde \mu}\prt_\tb X^{\tilde \nu}\eta_{\tilde \mu\tilde \nu}$,  by the pull-back metric \reef{tpull}. One also has to  write the partial derivatives of the gauge field strength in terms of the gauge field potential. The final resulting non-covariant expression involves independent structures  made of $F_{\ta\tb},\prt_\ta\prt_\tb A_\tc,\cdots$, $\prt_\ta X^{\tilde \mu},\prt_\ta \prt_\tb X^{\tilde \mu},\cdots $ and $\prt_\ta X^{y},\prt_\ta \prt_\tb X^{y},\cdots $. The coefficients of these independent structures which involve the parameters in the effective action found in the previous section, the parameters in the total derivative terms \reef{ttot} and the parameters in the field redefinitions \reef{tb40},  must be zero. They produces  some linear algebraic equations for these parameters. Solving them, one  finds some relations involving only the parameters of the independent couplings found in the previous section in which we are interested in this paper. The solution also produce some relations for $j_1,j_2,\cdots, j_{n_j}$, $k_1,k_2,\cdots, k_{n_k}$ in terms of the parameters of the effective action and  $j_{n_j+1},j_{n_j+2},\cdots$, $k_{n_k+1},k_{n_k+2},\cdots$ in which we are not interested.

Since the T-duality constraint \reef{tSJK} at order $\alpha'^2$ involve all orders of fields $F_{\ta\tb}, \prt_\ta X^y$, it relates the coefficients of all infinite number of independent couplings at order $\alpha'^2$. However, to solve this constraint one has to truncate the independent couplings in the effective action to a fixed number of $F,\Omega$. In the previous section we have found the couplings  up to six  $F,\Omega$.  To find  the parameters of these truncated  couplings, one has to truncate also the independent structures in \reef{tSJK}. If one considers an action at a given order of $\alpha'$, and at the level of $m$ fields $F,\Omega$, then the independent structures in the constraint \reef{tSJK} which have more than $m$ fields in the local frame must be ignored. The coefficients of the remaining independent structures must be zero. The resulting linear algebraic equation should be solved to find some relations between the parameters of the independent couplings in the action.

The independent couplings that we have found in the previous section have 12 couplings at the level of four $F,\Omega$, \ie the couplings with coefficients  $a_1,a_2,a_3,a_4,b_1,b_2,b_3,b_4,c_1,c_2,c_3,c_4$. To find the T-duality constraint on these couplings, we consider the following structures for the vector of the total derivative terms:
\beqa
{{\tilde \cI}} &\sim& \tilde D\tilde D{{\tilde \Omega }^y}\tilde D{X ^y}\tilde D{X ^y}\tilde D{X ^y} + {{\tilde \Omega }^y}\tilde D{{\tilde \Omega }^y}\tilde D{X ^y}\tilde D{X ^y} + {{\tilde \Omega }^y}{{\tilde \Omega }^y}{{\tilde \Omega }^y}\tilde D{X ^y}\nonumber\\
 &&+ F\tilde D\tilde D\tilde DF\tilde D{X ^y}\tilde D{X ^y} + \tilde D\tilde DF\tilde DF\tilde D{X ^y}\tilde D{X ^y} + F\tilde D\tilde DF{{\tilde \Omega }^y}\tilde D{X ^y}\nonumber\\
  &&+ \tilde DF\tilde DF{{\tilde \Omega }^y}\tilde D{X ^y} + FF{{\tilde \Omega }^y}\tilde D{{\tilde \Omega }^y} + F\tilde DF{{\tilde \Omega }^y}{{\tilde \Omega }^y} + F\tilde DF\tilde D{{\tilde \Omega }^y}\tilde D{X ^y}\nonumber\\
  &&+ FF\tilde D\tilde D{{\tilde \Omega }^y}\tilde D{X ^y} + \tilde D{X ^y}\tilde D{X ^y}{{\tilde \Omega }^{\tilde \mu }}\tilde D{{\tilde \Omega }^{\tilde \nu }} + \tilde D{X ^y}{{\tilde \Omega }^y}{{\tilde \Omega }^{\tilde \mu }}{{\tilde \Omega }^{\tilde \nu }}
\labell{a23}
\eeqa
For the field redefinitions, we consider the following structures:
\beqa
\delta {X ^y} &\sim& FF\tilde D\tilde D{{\tilde \Omega }^y} + F\tilde D\tilde D\tilde DF\tilde D{X ^y} + F\tilde DF\tilde D{{\tilde \Omega }^y}\nonumber\\
 &&+ F\tilde D\tilde DF{{\tilde \Omega }^y} + \tilde DF\tilde D\tilde DF\tilde D{X ^y} + \tilde DF\tilde DF{{\tilde \Omega }^y}\nonumber\\
 &&+ \tilde D\tilde D{{\tilde \Omega }^y}\tilde D{X ^y}\tilde D{X ^y} + {{\tilde \Omega }^y}\tilde D{{\tilde \Omega }^y}\tilde D{X ^y} + {{\tilde \Omega }^y}{{\tilde \Omega }^y}{{\tilde \Omega }^y}\nonumber\\
 &&+ {{\tilde \Omega }^{\tilde \mu }}\tilde D{{\tilde \Omega }^{\tilde \nu }}\tilde D{X ^y} + {{\tilde \Omega }^{\tilde \mu }}{{\tilde \Omega }^{\tilde \nu }}{{\tilde \Omega }^y}  \nn\\
\delta {A^{\tilde a}} &\sim& \tilde D\tilde D\tilde DF\tilde D{X ^y}\tilde D{X ^y} + \tilde D\tilde D{{\tilde \Omega }^y}\tilde D{X ^y}F + {{\tilde \Omega }^y}\tilde D{X ^y}\tilde D\tilde DF\nonumber\\
 &&+ \tilde D{{\tilde \Omega }^y}\tilde D{X ^y}\tilde DF + {{\tilde \Omega }^y}\tilde D{{\tilde \Omega }^y}F + {{\tilde \Omega }^y}{{\tilde \Omega }^y}\tilde DF \nn\\
\delta {X ^{\tilde \mu }} &\sim & \tilde D\tilde D{{\tilde \Omega }^{\tilde \mu }}\tilde D{X ^y}\tilde D{X ^y} + {{\tilde \Omega }^y}\tilde D{{\tilde \Omega }^{\tilde \mu }}\tilde D{X ^y} + {{\tilde \Omega }^{\tilde \mu }}\tilde D{{\tilde \Omega }^y}\tilde D{X ^y}+ {{\tilde \Omega }^{\tilde \mu }}{{\tilde \Omega }^y}{{\tilde \Omega }^y}
 \labell{cons30}
\eeqa
Note that all terms in the reduction $\Delta\tilde S$ in the constraint \reef{tSJK} involve, among other things,  $\prt X^y$ and/or $\tilde \Omega^y$. Hence the total derivative terms and the field redefinitions must include these fields as well. Using the package xAct, one can construct all possible contractions  in \reef{a23} and \reef{cons30}. Then replacing them in \reef{tSJK}, going to the local frame to write the equation \reef{tSJK} in terms of the independent structures, and removing  the terms that have six and more fields, one finds the resulting  linear algebraic equations have the following solution that involves only the parameters of the effective action:
\beqa
&&\,{c_1} \to \frac{{{a_1}}}{4} + \frac{{{a_3} - {a_4}}}{{16}},\,\,\,\,\,{c_2} \to  - \frac{3}{8}{a_3} + \frac{{{a_4}}}{4},\,\,\,\,\,{c_3} \to \frac{{{a_2} - {a_3} + {a_4}}}{4},\,\,\,\,\,{c_4} \to \frac{{{a_3}}}{2},\nonumber\\
&&\,{b_2} \to  - {a_2} +2 \,{b_1},\,\,\,\,\,{b_3} \to \frac{{{a_2} + {a_4}}}{4} - {b_1},\,\,\,\,\,{b_4} \to \frac{1}{4}(4{a_1} + {a_2} - 2{b_1})
 \labell{a600}
\eeqa
The above parameters are consistent with the results from the S-matrix method \reef{b4}. It turns out that the unfixed  parameters above can not be fixed by studying the T-duality constraint at the level of six $F,\Omega$. So for studying the constraint \reef{tSJK} at level of six $F,\Omega$, we consider the parameters \reef{b4} for the four $F,\Omega$ couplings.

We have found the independent couplings at the level of six $F,\Omega$ in the previous section, \ie the couplings with coefficients  $f_1,f_2,\cdots,f_{18}$, $d_1,d_2,\cdots, d_{64}$ and $e_1,e_2,\cdots, e_{64}$. To find the T-duality constraint on these couplings, we consider the following structures for the vector of the total derivative terms:
\beqa
{\tilde \cI} &\sim&  \tilde D\tilde D{{\tilde \Omega }^y}\tilde D{X ^y}\tilde D{X ^y}\tilde D{X ^y}\tilde D{X ^y}\tilde D{X ^y} + \tilde D{{\tilde \Omega }^y}{{\tilde \Omega }^y}\tilde D{X ^y}\tilde D{X ^y}\tilde D{X ^y}\tilde D{X ^y}\nonumber\\
  &&+ {{\tilde \Omega }^y}{{\tilde \Omega }^y}{{\tilde \Omega }^y}\tilde D{X ^y}\tilde D{X ^y}\tilde D{X ^y} + F\tilde D\tilde D\tilde DF\tilde D{X ^y}\tilde D{X ^y}\tilde D{X ^y}\tilde D{X ^y}\nonumber\\
  &&+ FF\tilde D\tilde D{{\tilde \Omega }^y}\tilde D{X ^y}\tilde D{X ^y}\tilde D{X ^y} + \tilde D\tilde DF\tilde DF\tilde D{X ^y}\tilde D{X ^y}\tilde D{X ^y}\tilde D{X ^y}\nonumber\\
  &&+ F\tilde D\tilde DF{{\tilde \Omega }^y}\tilde D{X ^y}\tilde D{X ^y}\tilde D{X ^y} + F\tilde DF\tilde D{{\tilde \Omega }^y}\tilde D{X ^y}\tilde D{X ^y}\tilde D{X ^y}\nonumber\\
  &&+ FF{{\tilde \Omega }^y}\tilde D{{\tilde \Omega }^y}\tilde D{X ^y}\tilde D{X ^y} + \tilde DF\tilde DF{{\tilde \Omega }^y}\tilde D{X ^y}\tilde D{X ^y}\tilde D{X ^y}\nonumber\\
  &&+ F\tilde DF{{\tilde \Omega }^y}{{\tilde \Omega }^y}\tilde D{X ^y}\tilde D{X ^y} + FF{{\tilde \Omega }^y}{{\tilde \Omega }^y}{{\tilde \Omega }^y}\tilde D{X ^y}\nonumber\\
  &&+ FFF\tilde D\tilde D\tilde DF\tilde D{X ^y}\tilde D{X ^y} + FFFF\tilde D\tilde D{{\tilde \Omega }^y}\tilde D{X ^y}\nonumber\\
 && + FF\tilde D\tilde DF\tilde DF\tilde D{X ^y}\tilde D{X ^y} + FFF\tilde D\tilde DF{{\tilde \Omega }^y}\tilde D{X ^y}\nonumber\\
  &&+ F\tilde DF\tilde DF\tilde DF\tilde D{X ^y}\tilde D{X ^y} + FF\tilde DF\tilde DF{{\tilde \Omega }^y}\tilde D{X ^y}\nonumber\\
  &&+ FFFF{{\tilde \Omega }^y}\tilde D{{\tilde \Omega }^y} + FFFF\tilde DF\tilde D{{\tilde \Omega }^y}\tilde D{X ^y}\nonumber\\
  &&+ FFF\tilde DF{{\tilde \Omega }^y}{{\tilde \Omega }^y} + {{\tilde \Omega }^{\tilde \mu }}\tilde D{{\tilde \Omega }^{\tilde \nu }}\tilde D{X ^y}\tilde D{X ^y}\tilde D{X ^y}\tilde D{X ^y}\nonumber\\
  &&+ {{\tilde \Omega }^{\tilde \mu }}{{\tilde \Omega }^{\tilde \nu }}{{\tilde \Omega }^y}\tilde D{X ^y}\tilde D{X ^y}\tilde D{X ^y} + FF{{\tilde \Omega }^{\tilde \mu }}\tilde D{{\tilde \Omega }^{\tilde \nu }}\tilde D{X ^y}\tilde D{X ^y}\nonumber\\
 && + F\tilde DF{{\tilde \Omega }^{\tilde \mu }}{{\tilde \Omega }^{\tilde \nu }}\tilde D{X ^y}\tilde D{X ^y} + FF{{\tilde \Omega }^{\tilde \mu }}{{\tilde \Omega }^{\tilde \nu }}{{\tilde \Omega }^y}\tilde D{X ^y}
\labell{a23a}
\eeqa
For the field redefinitions, we consider the following structures:
\beqa
\delta {X ^y} &\sim&  \tilde D\tilde D{{\tilde \Omega }^y}\tilde D{X ^y}\tilde D{X ^y}\tilde D{X ^y}\tilde D{X ^y} + {{\tilde \Omega }^y}\tilde D{{\tilde \Omega }^y}\tilde D{X ^y}\tilde D{X ^y}\tilde D{X ^y} + {{\tilde \Omega }^y}{{\tilde \Omega }^y}{{\tilde \Omega }^y}\tilde D{X ^y}\tilde D{X ^y}\nonumber\\
 &&+ \tilde D\tilde D{{\tilde \Omega }^y}FFFF + FFF\tilde D\tilde D\tilde DF\tilde D{X ^y} + \tilde D{{\tilde \Omega }^y}FFF\tilde DF\nonumber\\
 &&+ FF\tilde DF\tilde D\tilde DF\tilde D{X ^y} + {{\tilde \Omega }^y}FFF\tilde D\tilde DF + {{\tilde \Omega }^y}FF\tilde DF\tilde DF\nonumber\\
 &&+ F\tilde DF\tilde DF\tilde DF\tilde D{X ^y} + \tilde D\tilde D{{\tilde \Omega }^y}\tilde D{X ^y}\tilde D{X ^y}FF + F\tilde D\tilde D\tilde DF\tilde D{X ^y}\tilde D{X ^y}\tilde D{X ^y}\nonumber\\
 &&+ {{\tilde \Omega }^y}\tilde D{{\tilde \Omega }^y}\tilde D{X ^y}FF + \tilde D{{\tilde \Omega }^y}\tilde D{X ^y}\tilde D{X ^y}F\tilde DF + \tilde DF\tilde D\tilde DF\tilde D{X ^y}\tilde D{X ^y}\tilde D{X ^y}\nonumber\\
&& + {{\tilde \Omega }^y}F\tilde D\tilde DF\tilde D{X ^y}\tilde D{X ^y} + {{\tilde \Omega }^y}{{\tilde \Omega }^y}{{\tilde \Omega }^y}FF + {{\tilde \Omega }^y}{{\tilde \Omega }^y}F\tilde DF\tilde D{X ^y}\nonumber\\
&& + {{\tilde \Omega }^y}\tilde DF\tilde DF\tilde D{X ^y}\tilde D{X ^y} + {{\tilde \Omega }^{\tilde \mu }}\tilde D{{\tilde \Omega }^{\tilde \nu }}\tilde D{X ^y}\tilde D{X ^y}\tilde D{X ^y} + {{\tilde \Omega }^y}{{\tilde \Omega }^{\tilde \mu }}{{\tilde \Omega }^{\tilde \nu }}\tilde D{X ^y}\tilde D{X ^y}\nonumber\\
 &&+ FF{{\tilde \Omega }^{\tilde \mu }}\tilde D{{\tilde \Omega }^{\tilde \nu }}\tilde D{X ^y} + {{\tilde \Omega }^{\tilde \mu }}{{\tilde \Omega }^{\tilde \nu }}\tilde DF\tilde DF\tilde D{X ^y} + {{\tilde \Omega }^y}F\tilde DF{{\tilde \Omega }^{\tilde \mu }}{{\tilde \Omega }^{\tilde \nu }}\nn\\
\delta {A^{\tilde a}} &\sim&  FF\tilde D\tilde D\tilde DF\tilde D{X ^y}\tilde D{X ^y} + FFF\tilde D\tilde D{{\tilde \Omega }^y}\tilde D{X ^y}\nonumber\\
 &&+ F\tilde DF\tilde D\tilde DF\tilde D{X ^y}\tilde D{X ^y} + FF\tilde D\tilde DF{{\tilde \Omega }^y}\tilde D{X ^y} + FFF{{\tilde \Omega }^y}\tilde D{{\tilde \Omega }^y}\nonumber\\
 &&+ FF\tilde DF\tilde D{{\tilde \Omega }^y}\tilde D{X ^y} + \tilde DF\tilde DF\tilde DF\tilde D{X ^y}\tilde D{X ^y} + F\tilde DF\tilde DF{{\tilde \Omega }^y}\tilde D{X ^y}\nonumber\\
 &&+ FF\tilde DF{{\tilde \Omega }^y}{{\tilde \Omega }^y} + \tilde D\tilde D\tilde DF\tilde D{X ^y}\tilde D{X ^y}\tilde D{X ^y}\tilde D{X ^y} + F\tilde D\tilde D{{\tilde \Omega }^y}\tilde D{X ^y}\tilde D{X ^y}\tilde D{X ^y}\nonumber\\
 &&+ \tilde D\tilde DF{{\tilde \Omega }^y}\tilde D{X ^y}\tilde D{X ^y}\tilde D{X ^y} + \tilde DF\tilde D{{\tilde \Omega }^y}\tilde D{X ^y}\tilde D{X ^y}\tilde D{X ^y} + F{{\tilde \Omega }^y}\tilde D{{\tilde \Omega }^y}\tilde D{X ^y}\tilde D{X ^y}\nonumber\\
 &&+ \tilde DF{{\tilde \Omega }^y}{{\tilde \Omega }^y}\tilde D{X ^y}\tilde D{X ^y} + F{{\tilde \Omega }^y}{{\tilde \Omega }^y}{{\tilde \Omega }^y}\tilde D{X ^y} + F{{\tilde \Omega }^{\tilde \mu }}\tilde D{{\tilde \Omega }^{\tilde \nu }}\tilde D{X ^y}\tilde D{X ^y}\nonumber\\
 &&+ F{{\tilde \Omega }^{\tilde \mu }}{{\tilde \Omega }^{\tilde \nu }}{{\tilde \Omega }^y}\tilde D{X ^y} + \tilde DF{{\tilde \Omega }^{\tilde \mu }}{{\tilde \Omega }^{\tilde \nu }}\tilde D{X ^y}\tilde D{X ^y} \nn\\
\delta {X ^{\tilde \mu }} &\sim& \tilde D\tilde D{{\tilde \Omega }^{\tilde \mu }}\tilde D{X ^y}\tilde D{X ^y}\tilde D{X ^y}\tilde D{X ^y} + \tilde D{{\tilde \Omega }^{\tilde \mu }}{{\tilde \Omega }^y}\tilde D{X ^y}\tilde D{X ^y}\tilde D{X ^y}\nonumber\\
 &&+ {{\tilde \Omega }^{\tilde \mu }}\tilde D{{\tilde \Omega }^y}\tilde D{X ^y}\tilde D{X ^y}\tilde D{X ^y} + {{\tilde \Omega }^{\tilde \mu }}{{\tilde \Omega }^y}{{\tilde \Omega }^y}\tilde D{X ^y}\tilde D{X ^y} + \tilde D\tilde D{{\tilde \Omega }^{\tilde \mu }}\tilde D{X ^y}\tilde D{X ^y}FF\nonumber\\
 &&+ \tilde D{{\tilde \Omega }^{\tilde \mu }}{{\tilde \Omega }^y}\tilde D{X ^y}FF + \tilde D{{\tilde \Omega }^{\tilde \mu }}\tilde D{X ^y}\tilde D{X ^y}F\tilde DF + {{\tilde \Omega }^{\tilde \mu }}\tilde D{{\tilde \Omega }^y}\tilde D{X ^y}FF\nonumber\\
 &&+ {{\tilde \Omega }^{\tilde \mu }}\tilde D{X ^y}\tilde D{X ^y}F\tilde D\tilde DF + {{\tilde \Omega }^{\tilde \mu }}{{\tilde \Omega }^y}{{\tilde \Omega }^y}FF + {{\tilde \Omega }^{\tilde \mu }}\tilde D{X ^y}\tilde D{X ^y}\tilde DF\tilde DF\nonumber\\
 &&+ {{\tilde \Omega }^{\tilde \mu }}{{\tilde \Omega }^y}\tilde D{X ^y}F\tilde DF + {{\tilde \Omega }^{\tilde \mu }}{{\tilde \Omega }^{\tilde \nu }}{{\tilde \Omega }^y}\tilde D{X ^y}\tilde D{X ^y}
 \labell{cons300}
\eeqa
Using the package xAct, one can construct all possible contractions  in \reef{a23a} and \reef{cons300}. Then replacing them in \reef{tSJK}, using the parameters \reef{b4} for the four $F,\Omega$ couplings, going to the local frame to write the equation \reef{tSJK} in terms of the independent structures, and removing  the terms that have eight and more fields,   one finds the resulting  linear algebraic equations have the following solution that involves only the 146 parameters of the effective action:
\beqa
&&{d_2} \to \frac{1}{4}(8 + 6{d_1} - {d_{16}}),\,\,\,\,\,{d_3} \to \frac{5}{2} - 2{d_1},\,\,\,\,\,{d_4} \to  - \frac{{17}}{4} - {d_1} - {d_{16}},\,\,\,\,\,{d_5} \to 1 - 4{d_1},\nonumber\\
&&{d_6} \to  - 1 - {d_1},\,\,\,\,\,{d_7} \to  - 8,\,\,\,\,\,{d_8} \to \frac{3}{2} + 2{d_1},\,\,\,\,\,{d_9} \to  - 4,\,\,\,\,\,{d_{10}} \to  - 1 - 4{d_1},\nonumber\\
&&{d_{11}} \to 1 + 4{d_1},\,\,\,\,\,{d_{12}} \to 4,\,\,\,\,\,{d_{13}} \to 1 + 4{d_1},\,\,\,\,\,{d_{14}} \to  - 1 - 4{d_1},\,\,\,\,\,{d_{15}} \to  - 1,\nonumber\\
&&{d_{17}} \to 1 + 4{d_1},\,\,\,\,\,{d_{18}} \to  - \frac{1}{2} - 2{d_1},\,\,\,\,\,{d_{19}} \to 1 + 2{d_1},\,\,\,\,\,{d_{20}} \to \frac{{2 - {d_{16}}}}{4},\nonumber\\
&&{d_{21}} \to  - 1 + 4{d_1} + 2{d_{16}},\,\,\,\,\,{d_{22}} \to \frac{{27}}{2} + 6{d_1} - 5{d_{16}},\,\,\,\,\,{d_{23}} \to  - \frac{3}{2} - 2{d_1},\nonumber\\
&&{d_{24}} \to 7 + 4{d_1},\,\,\,\,\,{d_{25}} \to 13 + 8{d_1} - 2{d_{16}},\,\,\,\,\,{d_{26}} \to  - 8,\,\,\,\,\,{d_{27}} \to  - 21 - 12{d_1} + 4{d_{16}},\nonumber\\
&&{d_{28}} \to  - 7 - 4{d_1} + 2{d_{16}},\,\,\,\,\,{d_{29}} \to \frac{3}{2} + 2{d_1},\,\,\,\,\,{d_{30}} \to 2(6 + 6{d_1} - {d_{16}}),\nonumber\\
&&{d_{31}} \to  - 4(1 + {d_1}),\,\,\,\,\,{d_{32}} \to 2( - 2 + {d_{16}}),\,\,\,\,\,{d_{33}} \to 1,\,\,\,\,\,{d_{34}} \to 1 - 4{d_1},\nonumber\\
&&{d_{35}} \to  - 13 - 4{d_1} + 4{d_{16}},\,\,\,\,\,\,{d_{36}} \to \frac{1}{2} - 6{d_1} - {d_{16}},\,\,\,\,\,{d_{37}} \to \frac{{17}}{4} + 6{d_1} - \frac{{{d_{16}}}}{2},\nonumber\\
&&{d_{38}} \to  - 9 - 20{d_1},\,\,\,\,\,{d_{39}} \to 4 + 12{d_1},\,\,\,\,\,{d_{40}} \to 2 + 4{d_1} - \frac{{{d_{16}}}}{2},\,\nonumber\\
&&{d_{41}} \to  - 17 - 4{d_1} + 6{d_{16}},\,\,\,\,\,{d_{42}} \to 5 + 4{d_1},\,\,\,\,\,{d_{43}} \to 5 + 4{d_1},\,\,\,\,\,{d_{44}} \to 6 + 8{d_1} - 2{d_{16}},\nonumber\\
&&{d_{45}} \to 0,\,\,\,\,\,\,{d_{46}} \to  - 4 - 8{d_1},\,\,\,\,\,{d_{47}} \to 0,\,\,\,\,\,{d_{48}} \to 2 + 4{d_1},\,\,\,\,\,{d_{49}} \to  - \frac{3}{2},\nonumber\\
&&{d_{50}} \to  - \frac{5}{4} - {d_1},\,\,\,\,\,\,{d_{51}} \to \frac{5}{8} + \frac{3}{2}{d_1},\,\,\,\,\,{d_{52}} \to  - \frac{5}{{16}} - \frac{3}{4}{d_1},\,\,\,\,\,{d_{53}} \to  - \frac{5}{2},\nonumber\\
&&{d_{54}} \to \frac{5}{4} + {d_1},\,\,\,\,\,{d_{55}} \to  - \frac{{{d_1}}}{2},\,\,\,\,\,{d_{56}} \to 1,\,\,\,\,\,{d_{57}} \to 0,\,\,\,\,\,\,{d_{58}} \to 5 + 4{d_1},\nonumber\\
&&{d_{59}} \to  - 6 - 4{d_1} + 2{d_{16}},\,\,\,\,\,{d_{61}} \to  - \frac{1}{2} - 2{d_1} - {d_{16}} + {d_{60}},\,\,\,\,\,{d_{62}} \to  - 1,\nonumber\\
&&{d_{63}} \to  - 2 + {d_{16}},\,\,\,\,\,{d_{64}} \to 0,\nonumber\\
&&{e_1} \to \frac{1}{{16}}( - 7 + 4{d_1} + 2{d_{16}} + 2{d_{60}}),\,\,\,\,\,{e_2} \to \frac{3}{8}(1 + 4{d_1} - 2{d_{16}} + 2{d_{60}}),\nonumber\\
&&{e_3} \to \frac{1}{8}(7 - 4{d_1} - 2{d_{16}} - 2{d_{60}}),\,\,\,\,\,{e_4} \to  - {d_1} + \frac{5}{4}( - 3 + {d_{16}}),\nonumber\\
&&{e_5} \to \frac{1}{8}( - 7 + 4{d_1} + 2{d_{16}} + 2{d_{60}}),\,\,\,\,\,{e_6} \to \frac{1}{{16}}(19 + 12{d_1} - 14{d_{16}} + 6{d_{60}}),\nonumber\\
&&{e_7} \to 0,\,\,\,\,\,{e_8} \to 0,\,\,\,\,\,{e_9} \to \frac{1}{4}( - 13 +  - 4{d_1} + 8{d_{16}} - 2{d_{60}}),\nonumber\\
&&{e_{10}} \to \frac{1}{4}( - 7 + 4{d_1} + 2{d_{16}} + 2{d_{60}}),\,\,\,\,\,{e_{11}} \to 2( - 2 + {d_{16}}),\nonumber\\
&&{e_{12}} \to  - 2 + {d_{16}},\,\,\,\,\,{e_{13}} \to 0,\,\,\,\,\,{e_{14}} \to \frac{1}{8}( - 67 - 12{d_1} + 38{d_{16}} - 6{d_{60}}),\,\,\,\,\,{e_{15}} \to 0,\nonumber\\
&&{e_{16}} \to {d_1} + \frac{1}{4}( - 23 + 10{d_{16}} + 2{d_{60}}),\,\,\,\,\,{e_{17}} \to \frac{1}{8}( - 15 + 4{d_1} + 6{d_{16}} + 2{d_{60}}),\nonumber\\
&&{e_{18}} \to  - 6( - 2 + {d_{16}}),\,\,\,\,\,{e_{19}} \to \frac{1}{4}( - 9 - 4{d_1} + 6{d_{16}} - 2{d_{60}}),\nonumber\\
&&{e_{20}} \to \frac{1}{8}(1 + 4{d_1} - 2{d_{16}} + 2{d_{60}}),\,\,\,\,\,{e_{21}} \to 2( - 2 + {d_{16}}),\nonumber\\
&&{e_{22}} \to \frac{1}{8}(17 + 4{d_1} - 2{d_{16}} + 2{d_{60}}),\,\,\,\,{e_{23}} \to 2( - 2 + {d_{16}}),\,\,\,\,\,{e_{24}} \to 0,\nonumber\\
&&{e_{25}} \to \frac{1}{8}(15 - 4{d_1} - 6{d_{16}} - 2{d_{60}}),\,\,\,\,\,{e_{26}} \to \frac{1}{8}( - 49 - 4{d_1} + 10{d_{16}} - 2{d_{60}}),\nonumber\\
&&{e_{27}} \to 6 - {d_{16}},\,\,\,\,\,{e_{28}} \to 2( - 2 + {d_{16}}),\,\,\,\,\,{e_{29}} \to \frac{1}{{128}}(1 + 4{d_1} - 2{d_{16}} + 2{d_{60}}),\nonumber\\
&&{e_{30}} \to \frac{1}{8}(9 + 4{d_1} - 6{d_{16}} + 2{d_{60}}),\,\,\,\,\,{e_{31}} \to \frac{1}{{16}}(1 + 4{d_1} - 2{d_{16}} + 2{d_{60}}),\nonumber\\
&&{e_{32}} \to \frac{1}{2} + 2{d_1} - {d_{16}} + {d_{60}},\,\,\,\,\,{e_{33}} \to \frac{1}{8}( - 11 + 20{d_1} - 2{d_{16}} + 10{d_{60}}),\nonumber\\
&&{e_{34}} \to \frac{1}{8}(15 - 4{d_1} - 6{d_{16}} - 2{d_{60}}),\,\,\,\,\,{e_{35}} \to \frac{1}{2}( - 2 + {d_{16}}),\nonumber\\
&&{e_{36}} \to \frac{5}{4} + {d_1} - {d_{16}} + \frac{{{d_{60}}}}{2},\,\,\,\,\,{e_{37}} \to \frac{1}{{16}}(33 + 4{d_1} - 18{d_{16}} + 2{d_{60}}),\nonumber\\
&&{e_{38}} \to \frac{1}{8}( - 17 - 4{d_1} + 10{d_{16}} - 2{d_{60}}),\,\,\,\,\,{e_{39}} \to \frac{1}{8}( - 33 - 4{d_1} + 18{d_{16}} - 2{d_{60}}),\nonumber\\
&&{e_{40}} \to  - 2 + {d_{16}},\,\,\,\,\,{e_{41}} \to \frac{1}{8}( - 149 - 20{d_1} + 34{d_{16}} - 10{d_{60}}),\,\nonumber\\
&&{e_{42}} \to {d_1} + \frac{1}{4}(41 - 10{d_{16}} + 2{d_{60}}),\,\,\,\,\,{e_{43}} \to \frac{1}{8}( - 1 - 4{d_1} + 2{d_{16}} - 2{d_{60}}),\,\nonumber\\
&&{e_{44}} \to \frac{1}{8}(83 + 12{d_1} - 14{d_{16}} + 6{d_{60}}),\,\,\,\,\,{e_{45}} \to \frac{1}{8}( - 63 + 4{d_1} + 30{d_{16}} + 2{d_{60}}),\nonumber\\
&&{e_{46}} \to \frac{1}{{16}}(15 - 4{d_1} - 6{d_{16}} - 2{d_{60}}),\,\,\,\,\,{e_{47}} \to \frac{1}{8}(51 + 12{d_1} - 14{d_{16}} + 6{d_{60}}),\nonumber\\
&&{e_{48}} \to \frac{1}{4}( - 33 - 4{d_1} + 6{d_{16}} - 2{d_{60}}),\,\,\,\,\,{e_{49}} \to \frac{1}{{32}}( - 51 - 12{d_1} + 14{d_{16}} - 6{d_{60}}),\nonumber\\
&&{e_{50}} \to 0,\,\,\,\,\,{e_{51}} \to  - \frac{1}{2},\,\,\,\,\,{e_{52}} \to 0,\,\,\,\,\,{e_{53}} \to \frac{1}{8},\,\,\,\,\,{e_{54}} \to 0,\nonumber\\
&&{e_{55}} \to 0,\,\,\,\,\,{e_{56}} \to  - 2 + {d_{16}},\,\,\,\,\,{e_{57}} \to  - 2 + {d_{16}},\,\,\,\,\,{e_{58}} \to 0,\,\,\,\,\,{e_{59}} \to 0,\nonumber\\
&&{e_{60}} \to \frac{1}{8}( - 1 - 4{d_1} + 2{d_{16}} - 2{d_{60}}),\,\,\,\,\,{e_{61}} \to \frac{1}{8}(51 + 12{d_1} - 14{d_{16}} + 6{d_{60}}),\nonumber\\
&&{e_{62}} \to  - \frac{1}{2} - 2{d_1} + {d_{16}} - {d_{60}},\,\,\,\,\,{e_{63}} \to 1 - \frac{{{d_{16}}}}{2},\,\,\,\,\,{e_{64}} \to 0,\nonumber\\
&&{f_1} \to  - 4 + 3{d_{16}},\,\,\,\,\,{f_2} \to \frac{1}{2}(7 + 4{d_1} - {d_{16}}),\nonumber\\
&&{f_3} \to \frac{1}{4}(9 + 4{d_1} - 6{d_{16}} + 2{d_{60}}),\,\,\,\,\,{f_4} \to \frac{1}{4}(1 + 4{d_1} + 2{d_{16}} + 2{d_{60}}),\nonumber\\
&&{f_5} \to 4( - 2 + {d_{16}}),\,\,\,\,\,{f_6} \to 2,\,\,\,\,\,{f_7} \to 0,\,\,\,\,\,{f_8} \to  - 2,\,\,\,\,\,{f_9} \to 8 - 2{d_{16}},\nonumber\\
&&{f_{10}} \to  - 2,\,\,\,\,\,{f_{11}} \to \frac{1}{2},\,\,\,\,\,{f_{12}} \to  - 2,\,\,\,\,\,{f_{13}} \to  - 2 + {d_{16}},\,\,\,\,\,{f_{14}} \to 6,\nonumber\\
&&{f_{15}} \to  - \frac{1}{2},\,\,\,\,\,{f_{16}} \to  - 1,\,\,\,\,\,{f_{17}} \to  - 1,\,\,\,\,\,{f_{18}} \to 0.
 \labell{a700}
\eeqa
where the parameters, $d_1,d_{16}, d_{60}$ remain unfixed. We have also imposed the T-duality constraint \reef{tSJK} in the static gauge \cite{Karimi:2018vaf} and found exactly the same  result.

To check the above result, we compare them  with the correction at order $\alpha'^2$ to the Born-Infeld action that has been found by Wyllard  by the boundary state method \cite{Wyllard:2000qe}. This correction which involves all levels of the  gauge field strength is the following:
\beqa
S_{BI}^{(2)} =  - {T_p}\int {d^{p+1}\sigma \sqrt {-\det h_{ab}} [1 + \frac{{{{(2\pi \alpha ')}^2}}}{{96}}( - {h^{da}}{h^{bc}}{h^{fe}}{h^{uz}}{S_{euab}}{S_{zfcd}} + \frac{1}{2}{h^{fe}}{h^{uz}}{S_{eu}}{S_{zf}})]}
 \labell{a800}
\eeqa
where
\beqa
{S_{euab}} &=& {\partial _e}{\partial _u}{F_{ab}} + 2{h^{cd}}{\partial _e}{F_{[a|c}}{\partial _{u|}}{F_{b]d}}\nonumber\\
{S_{eu}} &=& {h^{ab}}{S_{euab}}\nn\\
 {h_{ab}} &=& {\eta _{ab}} + {F_{ab}}
 \labell{a900}
\eeqa
and  $h^{ab}$ is inverse of $h_{ab}$. The above action can be expanded to find the gauge field couplings at all levels of gauge field stength. The couplings at the level of four and six gauge field, should be related to the corresponding couplings in \reef{ac11} and \reef{ab11} with the parameters \reef{b4} and \reef{a700}, up to some total derivative terms and field redefinitions. Note that, since the parameters in the two actions are fixed, in order to compare the two actions, one should use all terms in the field redefintion $\cK$, \ie the dots in \reef{fred} should not be ignored. It has been verified in \cite{Wyllard:2000qe} that up to field redefinitions and total derivative terms, the two gauge field couplings are zero and the four gauge field couplings are the same as the corresponding couplings in \reef{ac11} with parameters \reef{b4}. We have compared the six gauge fields in the action \reef{a800} with the couplings in \reef{ab11} with the parameters in \reef{a700}. We have found they are the same up to some total derivative terms and field redefinitions provided that the unfixed parameter to be the  following:
\beqa
{d_1} = \frac{1}{4},\,\,\,\,\,{d_{16}} = 4,\,\,\,\,\,{d_{60}} =  - 1
 \labell{a9900}
\eeqa

To perform this calculation, we insert the six gauge field coupling of \reef{a800} on the right hand side of \reef{SS}, and the six independent gauge field couplings of \reef{ab11} with the parameters \reef{a700} on the left hand side of \reef{SS}. Then after imposing the Bianchi identity, one finds for some total derivative terms in $\cJ$ and field redefinitions in $\cK$, the two sets of couplings are  exactly the same if there is the above values for the unfixed parameters. This confirms that the  couplings  involving $F,\Omega$ that are fixed by the T-duality constraint \reef{tSJK} are consistent with the couplings involving $F$ that are fixed by the boundary state method.  Note that, neither the couplings in the  action \reef{a800} nor the independent couplings in our scheme,  include terms that have $D_aF^{ab}$. However, since the total derivative terms include terms that have $D_aF^{ab}$, in the comparison \reef{SS}, one must include the  field redefinition $\cK$.

Hence, the six gauge field strengths and/or the second fundamental forms are fixed in the  particular scheme that we have chosen in the previous section for the following parameters:
\beqa
&&{d_2} \to \frac{{11}}{8},\,\,\,\,\,{d_3} \to 2,\,\,\,\,\,{d_4} \to  - \frac{1}{2},\,\,\,\,\,{d_5} \to 0,\,\,\,\,\,{d_6} \to  - \frac{5}{4},\nonumber\\
&&
{d_7} \to  - 8,\,\,\,\,\,{d_8} \to 2,\,\,\,\,\,{d_9} \to  - 4,\,\,\,\,\,{d_{10}} \to  - 2,\,\,\,\,\,{d_{11}} \to 2,\nonumber\\
&&
{d_{12}} \to 4,\,\,\,\,\,{d_{13}} \to 2,\,\,\,\,\,{d_{14}} \to  - 2,\,\,\,\,\,{d_{15}} \to  - 1,\,\,\,\,\,{d_{17}} \to 2,\nonumber\\
&&
{d_{18}} \to  - 1\,\,\,\,\,{d_{19}} \to \frac{3}{2},\,\,\,\,\,{d_{20}} \to  - \frac{1}{2},\,\,\,\,\,{d_{21}} \to 8,\,\,\,\,\,{d_{22}} \to  - 5,\nonumber\\
&&
{d_{23}} \to  - 2,\,\,\,\,\,{d_{24}} \to 8,\,\,\,\,\,{d_{25}} \to 7,\,\,\,\,\,{d_{26}} \to  - 8,\,\,\,\,\,{d_{27}} \to  - 8,\nonumber\\
&&
{d_{28}} \to 0,\,\,\,\,\,{d_{29}} \to 2,\,\,\,\,\,{d_{30}} \to 7,\,\,\,\,\,{d_{31}} \to  - 5,\,\,\,\,\,{d_{32}} \to 4,\nonumber\\
&&
{d_{33}} \to 1,\,\,\,\,\,{d_{34}} \to 0,\,\,\,\,\,{d_{35}} \to 2,\,\,\,\,\,\,{d_{36}} \to  - 5,\,\,\,\,\,{d_{37}} \to \frac{{15}}{4},\nonumber\\
&&
{d_{38}} \to  - 14,\,\,\,\,\,{d_{39}} \to 7,\,\,\,\,\,{d_{40}} \to 1,\,\,\,\,\,\,{d_{41}} \to 6,\,\,\,\,\,{d_{42}} \to 6,\nonumber\\
&&
{d_{43}} \to 6,\,\,\,\,\,{d_{44}} \to 0,\,\,\,\,\,{d_{45}} \to 0,\,\,\,\,\,\,{d_{46}} \to  - 6,\,\,\,\,\,{d_{47}} \to 0,\nonumber\\
&&
{d_{48}} \to 3,\,\,\,\,\,{d_{49}} \to  - \frac{3}{2},\,\,\,\,\,{d_{50}} \to  - \frac{3}{2},\,\,\,\,\,\,{d_{51}} \to 1,\,\,\,\,\,{d_{52}} \to  - \frac{1}{2},\nonumber\\
&&
{d_{53}} \to  - \frac{5}{2},\,\,\,\,\,{d_{54}} \to \frac{3}{2},\,\,\,\,\,{d_{55}} \to  - \frac{1}{8},\,\,\,\,\,{d_{56}} \to 1,\,\,\,\,\,{d_{57}} \to 0,\nonumber\\
&&
{d_{58}} \to 6,\,\,\,\,\,{d_{59}} \to 1,\,\,\,\,\,{d_{61}} \to  - 6,\,\,\,\,\,{d_{62}} \to  - 1,\,\,\,\,\,{d_{63}} \to 2,\,\,\,\,\,{d_{64}} \to 0,\nonumber\\
&&
{e_1} \to 0,\,\,\,\,\,{e_2} \to  - 3,\,\,\,\,\,{e_3} \to 0,\,\,\,\,\,{e_4} \to 1,\,\,\,\,\,{e_5} \to 0,\nonumber\\
&&
{e_6} \to  - \frac{5}{2},\,\,\,\,\,{e_7} \to 0,\,\,\,\,\,{e_8} \to 0,\,\,\,\,\,{e_9} \to 5,\,\,\,\,\,{e_{10}} \to 0,\nonumber\\
&&
{e_{11}} \to 4,\,\,\,\,\,{e_{12}} \to 2,\,\,\,\,\,{e_{13}} \to 0,\,\,\,\,\,{e_{14}} \to 11,\,\,\,\,\,{e_{15}} \to 0,\nonumber\\
&&
{e_{16}} \to 4,\,\,\,\,\,{e_{17}} \to 1,\,\,\,\,\,{e_{18}} \to  - 12,\,\,\,\,\,{e_{19}} \to 4,\,\,\,\,\,{e_{20}} \to  - 1,\nonumber\\
&&
{e_{21}} \to 4,\,\,\,\,\,{e_{22}} \to 1,\,\,\,\,{e_{23}} \to 4,\,\,\,\,\,{e_{24}} \to 0,\,\,\,\,\,{e_{25}} \to  - 1,\nonumber\\
&&
{e_{26}} \to  - 1,\,\,\,\,\,{e_{27}} \to 2,\,\,\,\,\,{e_{28}} \to 4,\,\,\,\,\,{e_{29}} \to  - \frac{1}{{16}},\,\,\,\,\,{e_{30}} \to  - 2,\nonumber\\
&&
{e_{31}} \to  - \frac{1}{2},\,\,\,\,\,{e_{32}} \to  - 4,\,\,\,\,\,{e_{33}} \to  - 3,\,\,\,\,\,{e_{34}} \to  - 1,\,\,\,\,\,{e_{35}} \to 1,\nonumber\\
&&
{e_{36}} \to  - 3,\,\,\,\,\,{e_{37}} \to  - \frac{5}{2},\,\,\,\,\,{e_{38}} \to 3,\,\,\,\,\,{e_{39}} \to 5,\,\,\,\,\,{e_{40}} \to 2,\nonumber\\
&&
{e_{41}} \to  - 1,\,\,\,\,\,\,{e_{42}} \to 0,\,\,\,\,\,{e_{43}} \to 1,\,\,\,\,\,\,{e_{44}} \to 3,\,\,\,\,\,{e_{45}} \to 7,\nonumber\\
&&
{e_{46}} \to  - \frac{1}{2},\,\,\,\,\,{e_{47}} \to  - 1,\,\,\,\,\,{e_{48}} \to  - 2,\,\,\,\,\,{e_{49}} \to \frac{1}{4},\,\,\,\,\,{e_{50}} \to 0,\nonumber\\
&&
{e_{51}} \to  - \frac{1}{2},\,\,\,\,\,{e_{52}} \to 0,\,\,\,\,\,{e_{53}} \to \frac{1}{8},\,\,\,\,\,{e_{54}} \to 0,\,\,\,\,\,{e_{55}} \to 0,\nonumber\\
&&
{e_{56}} \to 2,\,\,\,\,\,{e_{57}} \to 2,\,\,\,\,\,{e_{58}} \to 0,\,\,\,\,\,{e_{59}} \to 0,\,\,\,\,\,{e_{60}} \to 1,\nonumber\\
&&
{e_{61}} \to  - 1,\,\,\,\,\,{e_{62}} \to 4,\,\,\,\,\,{e_{63}} \to  - 1,\,\,\,\,\,{e_{64}} \to 0,\nonumber\\
&&
{f_1} \to 8,\,\,\,\,\,{f_2} \to 2,\,\,\,\,\,{f_3} \to  - 4,\,\,\,\,\,{f_4} \to 2,\nonumber\\
&&
{f_5} \to 8,\,\,\,\,\,{f_6} \to 2,\,\,\,\,\,{f_7} \to 0,\,\,\,\,\,{f_8} \to  - 2,\,\,\,\,\,{f_9} \to 0,\nonumber\\
&&
{f_{10}} \to  - 2,\,\,\,\,\,{f_{11}} \to \frac{1}{2},\,\,\,\,\,{f_{12}} \to  - 2,\,\,\,\,\,{f_{13}} \to 2,\,\,\,\,\,{f_{14}} \to 6,\nonumber\\
&&
{f_{15}} \to  - \frac{1}{2},\,\,\,\,\,{f_{16}} \to  - 1,\,\,\,\,\,{f_{17}} \to  - 1,\,\,\,\,\,{f_{18}} \to 0.
 \labell{a1000}
\eeqa
The independent couplings at order $\alpha'^2$ in the previous section with the parameters \reef{b4} and \reef{a1000} include only four and six gauge field strengths and/or the second fundamental forms. It would be interesting to extend these couplings to all  levels of the gauge field strength. In the next section we study this extension.

\section{Towards all gauge field couplings}

 Using the  fact that the corrections to Born-Infeld action at order $\alpha'^2$ and at all levels of $F_{ab}$ are known \cite{Wyllard:2000qe}, one can easily extend these couplings to the covariant form by extending the partial derivatives in these couplings to the covariant derivatives and by extending the world volume flat metric to the pull-back of the bulk flat metric \reef{pull}, \ie
\beqa
 {S_p} &\supset &  - {T_{p}}\int {{d^{p + 1}}\sigma \sqrt {\det h_{ab}} [1 + \frac{{{{(2\pi \alpha ')}^2}}}{{96}}( - {h^{da}}{h^{bc}}{h^{fe}}{h^{uz}}{S_{euab}}{S_{zfcd}} + \frac{1}{2}{h^{fe}}{h^{uz}}{S_{eu}}{S_{zf}})]}
 \labell{a4400}
\eeqa
where
\beqa
{S_{euab}} &=& {D _e}{D_u}{F_{ab}} + 2{h^{cd}}{D _e}{F_{[a|c}}{D _{u|}}{F_{b]d}}\nonumber\\
{S_{eu}} &=& {h^{ab}}{S_{euab}}\nn\\
 {h_{ab}}& = &{{\tilde G}_{ab}} + {F_{ab}}
 \labell{a5500}
\eeqa
and  $ {h^{ab}}$ is inverse of $h_{ab}$.

One can expand the above action to find the four and six gauge field couplings with known coefficients. Then one can use them, and the couplings which involve $\Omega$ with the unknown coefficients that we have found them in a particular scheme in section 2, as the starting point for imposing the T-duality constraint \reef{tSJK}, to find the unknown coefficients. We have done this calculation and found the following couplings  for $\Omega\Omega\Omega\Omega$ and $\Omega\Omega\Omega\Omega FF$:
\begin{eqnarray}
  {S_p} &\supset & -\frac{(2\pi\alpha')^2}{96}{T_p}\int {{d^{p + 1}}\sigma \sqrt { - \det {{\tilde G}_{ab}}  } } \Big[ 2 \Omega _{a\,\,\,\,\mu}^{\,\,\,\,c}{\Omega ^{ab\mu}}\Omega _b^{\,\,\,\,d\nu}{\Omega _{cd\nu}}  -
  2\Omega _{ab}^{\,\,\,\,\,\,\nu}{\Omega ^{ab\mu}}{\Omega _{cd\nu}}\Omega _{\,\,\,\,\,\,\mu}^{cd} \nn\\&&+
   9\Omega _a^{\,\,\,\,e\mu}\Omega _b^{\,\,\,\,f\nu}{\Omega _{ce\nu}}{\Omega _{df\mu}}{F^{ab}}{F^{cd}}
   - 8\Omega _a^{\,\,\,\,e\mu}\Omega _{b\,\,\,\,\mu}^{\,\,\,\,f}\Omega _{ce}^{\,\,\,\,\,\,\nu}{\Omega _{df\nu}}{F^{ab}}{F^{cd}}
    \nn\\&&+3 \Omega _a^{\,\,\,\,e\mu}\Omega _{be}^{\,\,\,\,\,\,\nu}\Omega _{c\,\,\,\,\mu}^{\,\,\,\,f}{\Omega _{df\nu}}{F^{ab}}{F^{cd}}  + 8 \Omega _{ac}^{\,\,\,\,\,\,\mu}\Omega _b^{\,\,\,\,e\nu}\Omega _{d\,\,\,\,\nu}^{\,\,\,\,f}{\Omega _{ef\mu}}{F^{ab}}{F^{cd}}
     \nn\\&&+ 3 \Omega _b^{\,\,\,\,d\mu}\Omega _c^{\,\,\,\,e\nu}\Omega _{d\,\,\,\,\nu}^{\,\,\,\,f}{\Omega _{ef\mu}}F_a^{\,\,c}{F^{ab}}
  - 5 \Omega _b^{\,\,\,\,d\mu}\Omega _c^{\,\,\,\,e\nu}\Omega _{d\,\,\,\,\mu}^{\,\,\,\,f}{\Omega _{ef\nu}}F_a^{\,\,c}{F^{ab}}  \nn\\&&- 10 \Omega _{ac}^{\,\,\,\,\,\,\mu}\Omega _{b\,\,\,\,\mu}^{\,\,\,\,e}\Omega _d^{\,\,\,\,f\nu}{\Omega _{ef\nu}}{F^{ab}}{F^{cd}} \hfill
   - 3 \Omega _b^{\,\,\,\,d\mu}\Omega _{c\,\,\,\,\mu}^{\,\,\,\,e}\Omega _d^{\,\,\,\,f\nu}{\Omega _{ef\nu}}F_a^{\,\,c}{F^{ab}}  \nn\\&&+ \frac{1}{2} \Omega _{c\,\,\,\,\mu}^{\,\,\,\,e}{\Omega ^{cd\mu}}\Omega _d^{\,\,\,\,f\nu}{\Omega _{ef\nu}}{F_{ab}}{F^{ab}}
 -  {\Omega _{bc}}^\mu{\Omega _d}^{f\nu}{\Omega ^{de}}_\mu{\Omega _{ef\nu}} F_a^{\,\,c}{F^{ab}}  \nn\\&&+ 2 \Omega _{ac}^{\,\,\,\,\,\,\mu}\Omega _{bd}^{\,\,\,\,\,\,\nu}{\Omega _{ef\nu}}\Omega _{\,\,\,\,\,\,\mu}^{ef}{F^{ab}}{F^{cd}}
 + 5 \Omega _b^{\,\,\,\,d\mu}\Omega _{cd}^{\,\,\,\,\,\,\nu}{\Omega _{ef\nu}}\Omega _{\,\,\,\,\,\,\mu}^{ef}F_a^{\,\,c}{F^{ab}}
   \nn\\&&- \frac{1}{2} \Omega _{cd}^{\,\,\,\,\,\,\nu}{\Omega ^{cd\mu}}{\Omega _{ef\nu}}\Omega _{\,\,\,\,\,\,\mu}^{ef}{F_{ab}}{F^{ab}} - \frac{1}{2} \Omega _{ac}^{\,\,\,\,\,\,\mu}{\Omega _{bd\mu}}{\Omega _{ef\nu}}{\Omega ^{ef\nu}}{F^{ab}}{F^{cd}} \hfill
   \nn\\&&- \frac{1}{2} \Omega _b^{\,\,\,\,d\mu}{\Omega _{cd\mu}}{\Omega _{ef\nu}}{\Omega ^{ef\nu}}F_a^{\,\,c}{F^{ab}}  \hfill  \Big] \labell{a3000}
\end{eqnarray}
Note that the $\Omega\Omega\Omega\Omega$ terms above are exactly the second fundamental form correction that have been found in \cite{Bachas:1999um} up to terms that involve the trace of the second fundamental form, (see \cite{Wyllard:2001ye}), which are removed in our scheme.

We have also found the following couplings for the structures that include $DF$ or $D\Omega$:
\beqa
  {S_p} &\supset &  -\frac{(2\pi\alpha')^2}{96}{T_p}\int {{d^{p + 1}}\sigma \sqrt { - \det ({{\tilde G}_{ab}}  )} } \Big[\frac{1}{4}{D_a}{F_{bc}}{D^a}{F^{bc}}{\Omega _{de\mu}}{\Omega ^{de\mu}} + {D^a}{F^{bc}}{D^d}F_b^{\,\,e}\Omega _{ae}^{\,\,\,\,\,\,\mu}{\Omega _{cd\mu}}
    \nn\\&&- 2 {D^a}{F^{bc}}{D_b}F_a^{\,\,d}\Omega _c^{\,\,\,e\mu}{\Omega _{de\mu}}  - 5 {D^a}{F^{bc}}{D^d}F_b^{\,\,e}\Omega _{ac}^{\,\,\,\,\,\,\mu}{\Omega _{de\mu}}
   \nn\\&&-{D^a}{F^{bc}}{D^d}F_b^{\,\,e}{F_{fh}}{F^{fh}}\Omega _{ad}^{\,\,\,\,\,\,\mu}{\Omega _{ce\mu}}
   + \frac{3}{2}{D^a}{F^{bc}}{D_a}{F^{de}}{F_{fh}}{F^{fh}}\Omega _{bd}^{\,\,\,\,\,\,\mu}{\Omega _{ce\mu}}  \nn\\&&- {D^a}{F^{bc}}{D^d}{F^{ef}}F_b^{\,\,h}{F_{eh}}\Omega _{ad}^{\,\,\,\,\,\,\mu}{\Omega _{cf\mu}}
   - \frac{1}{2}{D^a}{F^{bc}}{D^d}{F^{ef}}F_a^{\,\,h}{F_{dh}}\Omega _{be}^{\,\,\,\,\,\,\mu}{\Omega _{cf\mu}}  \nn\\&&- 2 {D^a}{F^{bc}}{D_b}{F^{de}}F_d^{\,\,f}F_f^{\,\,h}\Omega _{ae}^{\,\,\,\,\,\,\mu}{\Omega _{ch\mu}}
  - 16 {D^a}{F^{bc}}{D^d}F_b^{\,\,e}F_a^{\,\,f}F_f^{\,\,h}\Omega _{ch}^{\,\,\,\,\,\,\mu}{\Omega _{de\mu}}
    \nn\\&&+ 5{D^a}{F^{bc}}{D^d}{F^{ef}}F_b^{\,\,h}{F_{eh}}\Omega _{ac}^{\,\,\,\,\,\,\mu}{\Omega _{df\mu}}  \hfill  + 4 {D^a}{F^{bc}}{D^d}{F^{ef}}F_a^{\,\,h}{F_{be}}\Omega _{ch}^{\,\,\,\,\,\,\mu}{\Omega _{df\mu}}
    \nn\\&&- 6 {D^a}{F^{bc}}{D^d}{F^{ef}}{F_{ae}}F_b^{\,\,h}\Omega _{ch}^{\,\,\,\,\,\,\mu}{\Omega _{df\mu}} - 2 {D^a}{F^{bc}}{D^d}{F^{ef}}{F_{ab}}F_e^{\,\,h}\Omega _{ch}^{\,\,\,\,\,\,\mu}{\Omega _{df\mu}} \nn\\&&- 4 {D^a}{F^{bc}}{D^d}{F^{ef}}F_a^{\,\,h}{F_{be}}\Omega _{cf}^{\,\,\,\,\,\,\mu}{\Omega _{dh\mu}}   + 6 {D^a}{F^{bc}}{D^d}{F^{ef}}{F_{ae}}F_b^{\,\,h}\Omega _{cf}^{\,\,\,\,\,\,\mu}{\Omega _{dh\mu}}
   \nn\\&&+2 {D^a}{F^{bc}}{D^d}{F^{ef}}{F_{ab}}F_e^{\,\,h}\Omega _{cf}^{\,\,\,\,\,\,\mu}{\Omega _{dh\mu}}  + 6 {D^a}{F^{bc}}{D_a}F_b^{\,\,d}F_e^{\,\,h}{F^{ef}}\Omega _{cf}^{\,\,\,\,\,\,\mu}{\Omega _{dh\mu}}
    \nn\\&&+ 4 {D^a}{F^{bc}}{D_b}F_a^{\,\,d}F_e^{\,\,h}{F^{ef}}\Omega _{cf}^{\,\,\,\,\,\,\mu}{\Omega _{dh\mu}}  \hfill - \frac{5}{2}{D^a}{F^{bc}}{D_a}F_b^{\,\,d}{F_{ef}}{F^{ef}}\Omega _c^{\,\,\,\,h\mu}{\Omega _{dh\mu}}
    \nn\\&&+3 {D^a}{F^{bc}}{D_b}F_a^{\,\,d}{F_{ef}}{F^{ef}}\Omega _c^{\,\,\,\,h\mu}{\Omega _{dh\mu}}  \hfill  + 10 {D^a}{F^{bc}}{D_b}{F^{de}}F_a^{\,\,f}F_d^{\,\,h}\Omega _{ch}^{\,\,\,\,\,\,\mu}{\Omega _{ef\mu}}
    \nn\\&&- 7 {D^a}{F^{bc}}{D^d}F_b^{\,\,e}F_a^{\,\,f}F_d^{\,\,h}\Omega _{ch}^{\,\,\,\,\,\,\mu}{\Omega _{ef\mu}} + 3 {D^a}{F^{bc}}{D_a}{F^{de}}F_b^{\,\,f}F_d^{\,\,h}\Omega _{ch}^{\,\,\,\,\,\,\mu}{\Omega _{ef\mu}}
    \nn\\&&+ 10 {D^a}{F^{bc}}{D_b}{F^{de}}F_d^{\,\,f}F_f^{\,\,h}\Omega _{ac}^{\,\,\,\,\,\,\mu}{\Omega _{eh\mu}}   + 4 {D^a}{F^{bc}}{D_b}{F^{de}}F_a^{\,\,f}F_f^{\,\,h}\Omega _{cd}^{\,\,\,\,\,\,\mu}{\Omega _{eh\mu}}
    \nn\\&&- 2 {D^a}{F^{bc}}{D_a}{F^{de}}F_b^{\,\,f}F_f^{\,\,h}\Omega _{cd}^{\,\,\,\,\,\,\mu}{\Omega _{eh\mu}}
   - 2 {D^a}{F^{bc}}{D_b}{F^{de}}F_a^{\,\,f}F_d^{\,\,h}\Omega _{cf}^{\,\,\,\,\,\,\mu}{\Omega _{eh\mu}}  \nn\\&&- 11 {D^a}{F^{bc}}{D_a}{F^{de}}F_b^{\,\,f}F_d^{\,\,h}\Omega _{cf}^{\,\,\,\,\,\,\mu}{\Omega _{eh\mu}}
   + 6 {D^a}{F^{bc}}{D_b}{F^{de}}{F_{ad}}{F^{fh}}\Omega _{cf}^{\,\,\,\,\,\,\mu}{\Omega _{eh\mu}}  \nn\\&&- 2 {D^a}{F^{bc}}{D_a}{F^{de}}{F_{bd}}{F^{fh}}\Omega _{cf}^{\,\,\,\,\,\,\mu}{\Omega _{eh\mu}}  - 2 {D^a}{F^{bc}}{D^d}F_b^{\,\,e}F_a^{\,\,f}{F_{df}}\Omega _c^{\,\,\,\,h\mu}{\Omega _{eh\mu}}  \nn\\&&+ 2 {D^a}{F^{bc}}{D^d}F_b^{\,\,e}F_a^{\,\,f}F_c^{\,\,h}\Omega _{df}^{\,\,\,\,\,\,\mu}{\Omega _{eh\mu}}
   - 5 {D^a}{F^{bc}}{D_a}{F^{de}}F_b^{\,\,f}F_c^{\,\,h}\Omega _{df}^{\,\,\,\,\,\,\mu}{\Omega _{eh\mu}}  \nn\\&&+ 4 {D^a}{F^{bc}}{D_a}{F^{de}}{F_{bc}}{F^{fh}}\Omega _{df}^{\,\,\,\,\,\,\mu}{\Omega _{eh\mu}}
   - 14 {D^a}{F^{bc}}{D_a}F_b^{\,\,d}F_c^{\,\,e}{F^{fh}}\Omega _{df}^{\,\,\,\,\,\,\mu}{\Omega _{eh\mu}}  \nn\\&&+ 6{D^a}{F^{bc}}{D_b}F_a^{\,\,d}F_c^{\,\,e}{F^{fh}}\Omega _{df}^{\,\,\,\,\,\,\mu}{\Omega _{eh\mu}}
   + 3{D^a}{F^{bc}}{D_b}{F_{ac}}{F^{de}}{F^{fh}}\Omega _{df}^{\,\,\,\,\,\,\mu}{\Omega _{eh\mu}} \nn\\&&+ 6{D^a}{F^{bc}}{D^d}F_b^{\,\,e}F_a^{\,\,f}F_d^{\,\,h}\Omega _{ce}^{\,\,\,\,\,\,\mu}{\Omega _{fh\mu}}  + 6 {D^a}{F^{bc}}{D^d}F_b^{\,\,e}F_a^{\,\,f}F_c^{\,\,h}\Omega _{de}^{\,\,\,\,\,\,\mu}{\Omega _{fh\mu}}  \nn\\&&- 4{D^a}{F^{bc}}{D^d}F_b^{\,\,e}{F_{ad}}F_c^{\,\,f}\Omega _e^{\,\,\,\,h\mu}{\Omega _{fh\mu}}
   + 4{D^a}{F^{bc}}{D_a}{F^{de}}{F_{bd}}F_c^{\,\,f}\Omega _e^{\,\,\,\,h\mu}{\Omega _{fh\mu}}  \nn\\&&+ 12{D^a}{F^{bc}}{D^d}F_b^{\,\,e}{F_{ac}}F_d^{\,\,f}\Omega _e^{\,\,\,\,h\mu}{\Omega _{fh\mu}}
   - 10 {D^a}{F^{bc}}{D_a}{F^{de}}{F_{bc}}F_d^{\,\,f}\Omega _e^{\,\,\,\,h\mu}{\Omega _{fh\mu}}  \nn\\&&- 4{D^a}{F^{bc}}{D_b}{F_{ac}}F_d^{\,\,f}{F^{de}}\Omega _e^{\,\,\,\,h\mu}{\Omega _{fh\mu}}
   + \frac{5}{2}{D^a}{F^{bc}}{D^d}F_b^{\,\,e}{F_{ad}}{F_{ce}}{\Omega _{fh\mu}}{\Omega ^{fh\mu}}  \nn\\&&- \frac{11}{4}{D^a}{F^{bc}}{D_a}{F^{de}}{F_{bd}}{F_{ce}}{\Omega _{fh\mu}}{\Omega ^{fh\mu}}
   + \frac{11}{8}{D^a}{F^{bc}}{D_a}{F^{de}}{F_{bc}}{F_{de}}{\Omega _{fh\mu}}{\Omega ^{fh\mu}}  \nn\\&&+ 2{D^a}{F^{bc}}{D_a}F_b^{\,\,d}F_c^{\,\,e}{F_{de}}{\Omega _{fh\mu}}{\Omega ^{fh\mu}}
   -  \frac{5}{2} {D^a}{F^{bc}}{D_b}F_a^{\,\,d}F_c^{\,\,e}{F_{de}}{\Omega _{fh\mu}}{\Omega ^{fh\mu}}  \nn\\&&+  \frac{3}{4}{D^a}{F^{bc}}{D_b}{F_{ac}}{F_{de}}{F^{de}}{\Omega _{fh\mu}}{\Omega ^{fh\mu}}
- 6 {D^a}{F^{bc}}{D_a}F_b^{\,\,d}F_e^{\,\,h}{F^{ef}}\Omega _{cd}^{\,\,\,\,\,\,\,\mu}{\Omega _{fh\mu}}  \nn\\&&+ 8{D^a}{F^{bc}}{D_a}{F^{de}}F_b^{\,\,f}F_d^{\,\,h}\Omega _{ce}^{\,\,\,\,\,\,\,\mu}{\Omega _{fh\mu}}
 + 2{D^a}{F^{bc}}{D^d}{F^{ef}}{F_{ab}}F_c^{\,\,h}\Omega _{de}^{\,\,\,\,\,\,\,\mu}{\Omega _{fh\mu}}  \nn\\&&- 10{D^a}{F^{bc}}{D^d}{F^{ef}}F_a^{\,\,h}{F_{bh}}\Omega _{ce}^{\,\,\,\,\,\,\,\mu}{\Omega _{df\mu}}
 + {D^a}{F^{bc}}{D^d}F_b^{\,\,e}F_a^{\,\,f}F_d^{\,\,h}\Omega _{cf}^{\,\,\,\,\,\,\,\mu}{\Omega _{eh\mu}}  \nn\\&&- 8{D^a}{F^{bc}}{D^d}F_b^{\,\,e}{F_{ad}}{F^{fh}}\Omega _{cf}^{\,\,\,\,\,\,\mu}{\Omega _{eh\mu}}
 + 8{D^a}{F^{bc}}{D^d}F_b^{\,\,e}F_a^{\,\,f}{F_{cf}}\Omega _d^{\,\,\,\,h\mu}{\Omega _{eh\mu}}  \nn\\&&+ 2{D^a}{\Omega ^{bc\mu}}{D^d}\Omega _{\,\,\,\,\,\,\mu}^{ef}{F_{ad}}{F_{be}}F_c^{\,\,h}{F_{fh}}
 - 4 {D^a}{F^{bc}}{D^d}{\Omega ^{ef\mu}}{F_{ab}}F_e^{\,\,h}{F_{fh}}{\Omega _{cd\mu}}
    \hfill  \Big] \labell{a2000}
\eeqa
One may extend the above calculations to find the covariant couplings involving $\Omega$ at the higher levels of gauge fields that are correspond to the action \reef{a4400}. It then rises the question that is it possible to find a compact expression for the covariant couplings involving $\Omega$ in terms of $h^{ab}$, as in \reef{a4400}? We have checked that the couplings in \reef{a3000} can not be written in terms of $\Omega\Omega\Omega\Omega hhhh$. The reason my be related to the particular scheme that we have used for the independent couplings in section 2. Even though the couplings with the structure $\Omega\Omega\Omega\Omega $ and $\Omega\Omega\Omega\Omega FF$ are independent of the scheme, but their coefficients that are fixed by the T-duality are scheme dependent because the T-duality relates these parameters to the parameters of the couplings involving $DF$ or $D\Omega$ which are scheme dependent.

\section{Conclusion}

In this paper we have found the independent world volume couplings at order $\alpha'^2$ involving four and six $F, \Omega$ and their covariant derivatives, in the normalization that $F$ is dimensionless. We have found that there are 12 couplings at four-field level and 146 couplings at the six-field level. The assumption that the effective action of the $D_p$-brane at the critical dimension is background independent is then used to find the parameters of the above independent couplings. That is, we have considered a particular background which has one circle. In this background, the effective action should satisfy the T-duality constraint \reef{tSJK}. This constraint fixes all parameters in terms of only 8 parameters. We have shown that these parameters are consistent with the all-gauge-field corrections to the Born-Infeld action that have been found in \cite{Wyllard:2000qe}. This comparison also fixes the remaining 8 parameters.  We have found the couplings in a particular scheme which is different from the scheme that has been used in \cite{Wyllard:2000qe}.

We then considered the couplings that have no second fundamental form to be the same as the couplings found by Wyllard in \cite{Wyllard:2000qe} in which the partial derivatives are extended to the covariant derivatives and the flat world volume metric to the pull-back of the bulk flat metric. We have found the covariant couplings involving four and six $F,\Omega$ that are consistent with these   couplings under the T-duality, \ie \reef{a3000} and \reef{a2000}. We could not succeed to extend them to all levels of $F$. The independent couplings \reef{a3000} and \reef{a2000} are in the  particular scheme  that we have considered in the section 2. That may be the reason  the covariant couplings \reef{a3000} and \reef{a2000} could not be written in a closed form in terms of $h^{ab}$ to include all levels of $F$.

To find the covariant couplings  at all levels of $F$, one may first need to find the independent covariant couplings involving $\Omega$, in terms of $h^{ab}$. That is, one should consider all gauge invariant couplings involving $DF,\Omega$ and their covariant derivatives at order $\alpha'^2$
\beqa
S'&=&-\frac{(2\pi\alpha')^2}{96} T_p\int d^{p+1}\sigma \, \sqrt { - \det h_{ab}}\,\cL'(DF,\cdots, \Omega, D\Omega,\cdots)\labell{s12}
\eeqa
where the spacetime index of $\Omega$ is contracted with the spacetime metric $\eta_{\mu\nu}$ and the world volume indices are contracted with $h^{ab}$. Then one adds to this action the total derivative terms and field redefinitions
\beqa
\cJ&=&-\alpha'^2 T_p\int d^{p+1}\sigma\, D_a\Bigg[\sqrt { - \det  h_{ab}}\, \cI^a\Bigg]\labell{tot2}\\
\cK&=&-\alpha'^2T_p\int d^{p+1}\sigma\,
 \sqrt { - \det h_{ab}}  \Bigg[
 -\frac{1}{2}(h^{ab}-h^{ba})D_a\delta {A_b}+ \frac{1}{2}(h^{ab}+h^{ba})\prt_a X^\mu D_b\delta {X^\nu }{\eta _{\mu \nu }}  \Bigg ]\nn
\eeqa
In this case, one has to use the identity \reef{Identity} to write the field redefinition terms produced by the last terms above, in terms of $\Omega$.   Doing the same steps as in section 2, one can find the independent couplings in terms of $h^{ab}$.
Then one may expand them and impose the T-duality constraint \reef{tSJK} to find the parameters of the independent couplings.  It would be interesting to find these covariant couplings in terms of $h^{ab}$, if they exist.  We have done this calculation for the covariant couplings at order $\alpha'$ in the bosonic string theory that have been found in \cite{Karimi:2018vaf} up to eight-field level, and found that there is no such covariant  couplings in terms of $h^{ab}$. It is in accord with the observation that  in the bosonic string theory, the world volume gravity couplings  on the $D_p$-brane in the presence of constant $B$-field, in terms of inverse of  $h_{ab}=\tG_{ab}+B_{ab}+F_{ab}$  can not be written in a covariant form   at order $\alpha'$ \cite{Ardalan:2002qt}.

The $D_p$-brane action in type II superstring theory has also the Wess-Zumino coupling that at the lowest order of $\alpha'$   involves the R-R potential and $F_{ab}$  \cite{DiVecchia:1999uf}. The $\alpha'^2$ corrections to this action  involving only $F$ and its covariant derivatives, have been found in \cite{Wyllard:2000qe}. The corrections that involve only $\Omega$ have been found in \cite{Bachas:1999um}. It would be interesting to use the T-duality constraint to find the correction  that involve $F,\Omega$ and their covariant derivatives at order $\alpha'^2$, as we have done in this paper for the DBI action.

We have used the field redefinitions to remove the couplings that have  $D_aF^{ab}$ or $G^{ab}{\Omega _{ab}}^\nu$. These couplings are not produced by the disc-level S-matrix elements of massless open string vertex operators either. However, the second fundamental form in non-trivial bulk background has gravity contribution as well as the transverse scalar contributions. If one uses the bulk field redefinitions to write the bulk effective action in a fixed scheme, then one would not be allowed  to use the field redefinitions to remove the trace of the second fundamental form from the world volume effective action of the D$_p$-brane. Hence, the world volume couplings involving the trace of the second fundamental form should be reproduced by the disc-level S-matrix elements of massless closed string vertex operators. In fact such couplings have been found in the  bosonic string theory at order $\alpha'$ in \cite{Corley:2001hg} and in the superstring theories at order $\alpha'^2$ in \cite{Bachas:1999um}. Similarly, the couplings involving $D_aF^{ab}$ have closed string contribution through the standard replacement of $F$ by $F+B$ in the world volume effective actions. If one uses the bulk field redefinitions to write the bulk effective action in a fixed scheme, then one would not be allowed  to use the field redefinitions to remove $D_aB^{ab}$ from the world volume effective action of the D$_p$-brane. Hence such couplings may be reproduced by the disc-level S-matrix elements of massless closed string vertex operators. Alternatively, one may  find the couplings involving $D_aB^{ab}$ and the second fundamental form by studying the T-duality constraint on the world volume couplings of massless closed string  states \cite{Hosseini}. 
\\

{\bf Acknowledgements}:  The research of  S. Karimi is supported by Ferdowsi University of Mashhad under grant  35315(1400/06/22).


\end{document}